\documentclass[%
 reprint,
 amsmath,amssymb,
 aps,
]{revtex4-2}

\usepackage{graphicx}
\usepackage{color}
\usepackage{dcolumn}
\usepackage{bm}

\begin{document}

\preprint{APS/123-QED}

\title{Development of Kinetic Energy Density Functional Using \\ Response Function Defined on the Energy Coordinate}

\author{Hideaki Takahashi}
 \email{ hideaki.takahashi.c4@tohoku.ac.jp}
\affiliation{%
Department of Chemistry, Graduate School of Science,\\
 Tohoku University, Sendai, Miyagi 980-8578, Japan 
}%




\date{\today}

\begin{abstract}
A kinetic energy functional $E_\text{kin}^e$ was developed within the framework of the density-functional theory (DFT) based on the energy electron density for the purpose of realizing the orbital-free DFT. The functional includes the nonlocal term described with the linear-response function (LRF) of a reference system. As a notable feature of the present approach, the LRF is represented on the energy coordinate $\epsilon$ defined for each system of interest. In addition, an atomic system is taken as a reference system for the construction of the LRF, which shows a clear difference from the conventional approach based on the homogeneous electron gas. The explicit form of the functional $E_\text{kin}^e$ was formulated by means of the coupling-parameter integration scheme. The functional $E_\text{kin}^e$ was applied to the calculations of the kinetic energies of the pseudo atoms that mimics H, He, Ne, and Ar. Explicitly, the kinetic energy of each atom was computed using the functional $E_\text{kin}^e$ with respect to the variation of the valence charge $Z_v$ of each atom. In these calculations, the electron density $n$ optimized by the Kohn-Sham DFT was adopted as an argument of the functional. It was found that the results are in excellent agreements with those given by the Kohn-Sham DFT. We also devised a method to perform the self-consistent field calculation utilizing the functional $E_\text{kin}^e$. The method was applied to the computation of the radial distribution functions of the electrons in the pseudo Ne and Ar atoms. It was demonstrated that the results reasonably agree with those yielded by the Kohn-Sham DFT.       
   
\end{abstract}

\maketitle


\section{\label{sec:level1}Introduction}

Density-functional theory (DFT) for electrons\cite{rf:parr_yang_eng} is one of the principal subjects in modern physics and chemistry. It offers an efficient and robust theoretical framework to study electronic structures of materials and molecules\cite{Martin2004}. Vast amounts of effort have been devoted to the  developments of functionals and to the foundation of the theoretical basis of DFT\cite{Engel2011}. Nowadays, DFT is an indispensable tool in the computational approach to the material design and molecular synthesis because of its efficiency and reliability. In the early study of DFT it was  proved in formulation\cite{Levy1979pnas, rf:parr_yang_eng} that there exists an exact universal functional $E[n]$ that describes the total energy $E$ in terms of the $N$-representable\cite{rf:parr_yang_eng, Levy1979pnas} electron density $n$.  However, the Kohn-Sham (KS) method\cite{rf:kohn1965pr, rf:parr_yang_eng}, which has so far been the only workhorse in the DFT, utilizes one-electron orbitals as the variables in the variational search. The introduction of the wave functions in KS-DFT is necessitated to ensure the accurate evaluation of the kinetic energy of the electrons. However, it gives rise to the computational cost associated with the orthogonalization among the orbitals, which roughly scales as $O$($N^3$) with the system size $N$.  If the construction of an effective kinetic energy functional is made possible, the one-electron orbitals become no longer necessary, and consequently, the computational cost scales linearly with the system size (\textquoteleft Order-$N$\textquoteright) since the orthogonalization can be completely bypassed in the variational calculations. Thus, it is a matter of great significance in physics and chemistry to develop the efficacious orbital-free DFT (OF-DFT) to extend the frontier of the applications of DFT\cite{Wesolowski2013}.   
 
Development of an accurate kinetic energy density functional $E_\text{kin}[n]$ is, however, known as the toughest subject. A primitive approximation to $E_\text{kin}[n]$ was first given in 1928 by Thomas and Fermi (TF)\cite{Thomas1927PCPS, Fermi1928zp} using the homogeneous electron gas (HEG)\cite{rf:parr_yang_eng} as a reference system. For the total electron density $n$, the TF functional $E_\text{TF}[n]$ has the form
\begin{equation}
E_\text{TF}[n] = C_\text{TF} \int d\bm{r}\; n(\bm{r})^{\frac{5}{3}}
\label{eq:E_TF}
\end{equation}
where $C_\text{TF}=\frac{3}{10}\left(3 \pi^{2}\right)^{\frac{2}{3}}$ and $\bm{r}$ is the position vector of the electrons. In 1935 a gradient correction $E_\text{vW}[n]$ 
\begin{equation}
E_\text{vW}[n] = -\frac{1}{2}  \int d\bm{r} \; n(\bm{r})^{1 / 2} \nabla^{2} n(\bm{r})^{1 / 2}
\label{eq:E_vW}
\end{equation}
was made by von Weizs\"{a}cker.\cite{Weizsacker1935zp} It is readily recognized that Eq. (\ref{eq:E_vW}) by itself is an exact $E_\text{kin}[n]$ for one or two electrons systems. Inclusion of $E_\text{vW}[n]$ with a prefactor $\lambda$ ($\lambda$vW)
\begin{equation}
E_\text{TFW}[n]= E_\text{TF}[n] + \lambda E_\text{vW}[n]
\label{eq:TF+lvW}
\end{equation}
makes a substantial contribution to reduce errors. The behavior of the linear response function(LRF) of $E_\text{TFW}[n]$ in the momentum space  was analyzed, for instance, in Ref. \onlinecite{Wang2000}. It shows that LRF of $E_\text{TFW}[n]$ with $\lambda = 1$ reproduces the asymptotic behavior of the HEG in the large $\bm{k}$ limit, while setting $\lambda=1/9$ realizes that in the small $\bm{k}$ region. As demonstrated in Ref. \onlinecite{Yonei1965jpsj}, $\lambda=1/5$ gives the best results when it is applied to atoms. Unfortunately, higher order corrections to $E_\text{TFW}[n]$ offers quite minor improvements at the most.  It has long been considered that the deficiency of these local and semilocal kinetic operators is that they would not be able to realize the shell structures of electron densities in atoms. This motivated the introduction of the nonlocal term in the functional $E_\text{kin}[n]$. 

The nonlocal effect was first incorporated into the kinetic functional $E_\text{CAT}$ by Chac\'{o}n, Alvarellos, and Tarazona\cite{Chacon1985prb} through the weighted average $\overline{n}(\bm{r})$ of the electron density $n$. The weight function was constructed so that the functional realizes the linear response of the HEG. Another type of the nonlocal kinetic functional $E_\text{WT}[n]$ was proposed by Wang and Teter\cite{Wang1992prb}, which also refers to the LRF of the HEG to formulate the integral kernel in the functional. Explicitly, the nonlocal term $E_\text{WT}^\text{nloc}[n]$ in the WT functional is in the form 
\begin{equation}
E_\text{WT}^\text{nloc}[n] = C_{\text{WT}}^\text{nloc} \int d \bm{r} d \bm{r}^{\prime} n(\bm{r})^\alpha \omega_0\left(k_F|\bm{r}-\bm{r}^{\prime}|\right) n\left(\bm{r}^{\prime}\right)^\alpha
\label{eq:E_WT}
\end{equation}
where $k_F$ is the Fermi momentum of the HEG with density $n_0$ and is given by $k_{F}=\left(3 \pi^{2} n_{0}\right)^{1 / 3}$. $\alpha$ in Eq. (\ref{eq:E_WT}) was taken as $5/6$ in the original development. For the bulk system, the average electron density in the unit cell can be adopted to $n_0$. The function $\omega_0$ is related to the LRF of the HEG and it has an oscillatory behavior in real space. It is, thus, plausible that the functional is capable of producing the quantum oscillations responsible for the creation of the shell structures in atoms. Actually, Figs. 8 and 9 in Ref. \onlinecite{Wang1992prb} show that the $E_\text{WT}[n]$ functional yields weak, but apparent shell structures in the density profiles of atoms. Motivated by this work, Wang, Govind, and Carter made an improvement on the WT functional in Ref. \onlinecite{Wang1998prb} and, further, developed a density-dependent kernel in Ref. \onlinecite{Wang1999prb}. However, adapting the LRF of the HEG to atomic or molecular systems might be problematic since it is not straightforward to specify the corresponding Fermi momentum $k_F$ for the given different positions $\bm{r}$ and $\bm{r}^\prime$ since the average electron density $n_0$ is not well defined in contrast to the bulk systems. Along this line, Huang and Carter devised a method which incorporates the gradient of the density into the kernel for the applications to semiconductors\cite{Huang2010prb}.  However, even when $k_F$ is determined somehow, there is no good reason that LRF for the HEG will be successfully applied to the systems with large inhomogeneities. It is desirable to build a novel LRF based on another reference system to construct the nonlocal kinetic energy functional suitable for the applications to atoms and molecules.

In a recent work\cite{Takahashi2018jpb}, a new DFT for electrons was developed on the basis of the electron distribution $n^e(\epsilon)$ on the energy coordinate $\epsilon$.  The distribution $n^e(\epsilon)$, referred to as energy electron density, is given by the projection of the density $n(\bm{r})$ onto the coordinate $\epsilon$, thus, 
\begin{equation}               
n^e(\epsilon) = \int d\bm{r}\; n(\bm{r}) \delta(\epsilon - \upsilon_\text{def}(\bm{r})) 
\label{eq:n_e}
\end{equation}
where $\upsilon_\text{def}(\bm{r})$ is the potential introduced to define the energy coordinate and the external potential $\upsilon_\text{ext}(\bm{r})$ of interest is adopted usually. Explicitly, the coordinate $\epsilon$ is given by
\begin{equation}
\upsilon_{\text {def }}(\bm{r})=\sum_\text{A} \frac{Z_\text{A}}{\left|\bm{r}-\bm{R}_\text{A}\right|}
\label{eq:epsilon}
\end{equation}
where $\bm{R}_\text{A}$ and $Z_\text{A}$ are the position vector and the charge of the nucleus $\text{A}$, respectively. Importantly, it can be proved in parallel to the Hohenberg-Kohn theorem\cite{rf:hohenberg1964pr} that there exists one-to-one correspondence between certain \textit{subsets} of the energy electron densities and the external potentials\cite{Takahashi2018jpb}. Furthermore, the Levy's constraint search method\cite{Levy1979pnas} can also be formulated within the DFT based on the energy electron density\cite{Takahashi2018jpb}. The introduction of the energy electron density $n^e(\epsilon)$ is motivated by the fact that it offers a natural framework to incorporate the nonlocal electron correlations which are crucial in describing the chemical bond dissociations. The correlation is often referred to as static correlation(SC) or left-right correlation. In Ref. \onlinecite{Takahashi2018jpb} a prototype of the SC energy functional was constructed utilizing $n^e(\epsilon)$ and a more sophisticated functional was developed in a subsequent paper\cite{Takahashi2020}.        

In the present work, a prototype of the kinetic energy functional is developed for the applications to atoms, where a nonlocal kinetic potential is constructed on the energy coordinate. It should also be noted that the LRF of the individual atomic system instead of HEG is utilized to build the nonlocal term in contrast to the previous works. At this stage, our kinetic energy functional is only applicable to atoms. The performance is examined by computing the energies of the atoms with shifted nuclear charges. The radial electron distributions around the nucleus for the atom are also computed. The results are compared with those obtained by using the functional $E_\text{TFW}[n]$ in Eq. (\ref{eq:TF+lvW}) and those by the KS-DFT calculation.     

The organization of the paper is as follows. In Section II, we provide the theoretical details for the construction of the nonlocal kinetic energy functional of the energy electron density, following a brief review of the LRF of the HEG is presented at first. Section III is devoted to describe the details of the numerical implementations and calculations. The results and discussion are presented in Section IV. The summary and the perspective are given in the last section.    

\section{\label{sec:level2}Theory and Method}
In this section, theoretical and methodological details are provided for the construction of the kinetic energy functional. In subsection A, we briefly review the relation of the functional derivative of the kinetic potential with the linear-response function of the system. Then, we illustrate the projection of the nonlocal term in the kinetic energy functional onto the energy coordinate with a focus placed on its justification. In the third subsection, the explicit form of the kinetic energy functional is developed on the basis of the coupling-parameter integration. The methodology to realize the self-consistent field calculation (SCF) with the present approach is given in the last subsection.    
\subsection{Linear-Response Function}
We first review the relationship between the kinetic potential and the linear response function(LRF) of the system of interest for later reference. We assume that an energy functional $E[n]$ of the density $n(\bm{r})$ is given.  In the density functional framework, $E[n]$ can be further decomposed into the contributions
\begin{equation}  
E[n] = E_\text{kin}[n]+E_\text{H}[n]+E_{xc}[n]+E_\text{ext}[n]
\label{eq:E_decomp}
\end{equation}
In Eq. (\ref{eq:E_decomp}), $E_\text{H}[n]$ is the Hartree energy due to the classical Coulomb repulsion among electrons, $E_{xc}[n]$ is the exchange-correlation energy, and $E_\text{ext}$ is the interaction energy between the density $n$ and the external potential $\upsilon_\text{ext}(\bm{r})$. At the equilibrium of the electronic state, the functional derivative of the energy with respect to the density becomes the chemical potential $\mu$ of the electron, 
\begin{align}  
\frac{\delta E[n]}{\delta n(\bm{r})} &=\frac{\delta E_\text{kin}[n]}{\delta n(\bm{r})}      \notag     \\
& \;\;\;\; +\frac{\delta}{\delta n(\bm{r})}\left(E_\text{H}[n]+E_{x c}[n]+E_{\text {ext }}[n]\right)        \notag    \\
& = \upsilon_\text{kin}[n](\bm{r}) + \upsilon_\text{eff}[n](\bm{r})  =\mu 
\label{eq:E_der1}
\end{align}
where $\upsilon_\text{eff}[n](\bm{r})$ is the effective potential due to the electron correlations and the external potential. Since it is assumed the system is in the stationary state, the second derivative of the left hand side of Eq. (\ref{eq:E_der1}) leads to zero, 
\begin{equation}
\frac{\delta^{2} E[n]}{\delta n(\bm{r}) \delta n\left(\bm{r}^{\prime}\right)}=\frac{\delta \upsilon_\text{kin}[n](\bm{r})}{\delta n\left(\bm{r}^{\prime}\right)}+\frac{\delta \upsilon_\text{eff}[n](\bm{r})}{\delta n\left(\bm{r}^{\prime}\right)}=0
\label{eq:E_der2}
\end{equation}  
It should be noticed that the second term of the right hand side of the first equality in Eq. (\ref{eq:E_der2}) represents the inverse of the response function $\chi(\bm{r}, \bm{r}^\prime)$ of the system. Thus, the functional derivative of the kinetic potential is directly related to the response function of the system, 
\begin{equation} 
\frac{\delta \upsilon_\text{kin}[n](\bm{r})}{\delta n\left(\bm{r}^{\prime}\right)} = -\frac{\delta \upsilon_\text{eff}[n](\bm{r})}{\delta n\left(\bm{r}^{\prime}\right)} = -\chi^{-1}\left(\bm{r}, \bm{r}^{\prime}\right)
\label{eq:vkin_der}
\end{equation}
This equation constitutes the basis of the constructions of the nonlocal kinetic functionals $E_\text{CAT}$\cite{Chacon1985prb}, $E_\text{WT}$\cite{Wang1992prb}, and their modifications. Importantly, all of these studies refer to the Lindhard function\cite{Lindhard1954}, that is, the linear-response function $\chi_0(k_F;|\bm{r}-\bm{r}^\prime|)$ of the homogeneous electron gas (HEG) for the sake of numerical convenience.   In the construction of the nonlocal energy in the WT functional\cite{Wang1992prb}, for instance, the integral kernel $\omega$ in Eq. (\ref{eq:E_WT}) is obtained by excluding the terms corresponding to the local and semi-local energies $E_\text{TF}$ and $E_\text{vW}$ from $-\chi_0^{-1}$.  Although the average density $n_0$ is well defined in bulk system, it cannot be determined for atoms and molecules at least in a natural manner. Moreover, the applicability of the LRF of the HEG to atoms and molecules that have extreme inhomogeneity is not yet well examined. It is, thus, desirable to devise a new approach using the LRF for some \textit{inhomogeneous} reference system to construct the kinetic energy density functional applicable to the systems with large inhomogeneities.   

\subsection{Projection of Nonlocal Kinetic Energy Functional onto the Energy Coordinate}
As mentioned in Introduction, a novel framework of the DFT was developed\cite{Takahashi2018jpb} in 2018 on the basis of the energy electron density $n^e(\epsilon)$ defined in Eq. (\ref{eq:n_e}). The construction of this theory was, in part, motivated by a hypothesis by Parr et al. in Ref. \onlinecite{rf:parr1981plenum}. They advocated that the contours of the electron densities of molecules are more or less parallel to those of the bare nuclear potentials, that is, Eq. (\ref{eq:epsilon}) itself. Actually, this is valid for the densities in atomic systems.  With this in mind, we reformulate the nonlocal term $E_\text{WT}^\text{nloc}[n]$ in Eq. (\ref{eq:E_WT}) with the method of the energy representation. First, we consider the general form $E_{\text{gWT}}^{\text{nloc}}[n]$ of the nonlocal term of the WT functional, which may be written as  
\begin{equation}  
E_{\text{gWT}}^{\text{nloc}}[n] = C_{\text{gWT}}^{\text{nloc}} \int d \bm{r} d \bm{r}^{\prime} n(\bm{r})^{\alpha} \omega\left(\bm{r},\bm{r}^{\prime}\right) n\left(\bm{r}^{\prime}\right)^{\alpha}
\label{eq:E_gWT}
\end{equation}
Note that the integral kernel $\omega_0$ built for the HEG is replaced by some general function $\omega$ in Eq. (\ref{eq:E_gWT}).  
We also suppose that the following relation is satisfied at least approximately,
\begin{equation}  
n(\bm{r}) = \left. \widetilde{n}^e(\epsilon) \right|_{ \epsilon=\upsilon_\text{def}(\bm{r})}
\label{eq:constancy}
\end{equation}
where $\widetilde{n}^e(\epsilon)$ is a certain function of $n^e(\epsilon)$. \textcolor{black}{ In Eq. (\ref{eq:constancy}), the density $n(\bm{r})$ is supposed to be nearly constant on the hypersurface with the energy coordinate $\epsilon$ specified by $\epsilon=\upsilon_\text{def}(\bm{r})$. $n(\bm{r})$ can be regarded as a composite function, thus, $n(\bm{r})=\widetilde{n}^e ( \upsilon_\text{def} (\bm{r}) )$. Eq. (\ref{eq:E_gWT}) can then be cast into the form of the energy representation, 
\begin{align}
E_\text{gWT}^{\text{nloc}}[n] = & C_{\text{gWT}}^{\text{nloc}} \int d \bm{r} d \bm{r}^{\prime} n(\bm{r})^{\alpha} \omega\left(\bm{r}, \bm{r}^{\prime}\right) n\left(\bm{r}^{\prime}\right)^{\alpha}    \notag  \\
= & C_{\text{gWT}}^{\text{nloc}} \int d \epsilon d \epsilon^{\prime} \int d \bm{r} d \bm{r}^{\prime} \delta(\epsilon-\upsilon_\text{def}(\bm{r}))  \notag    \\
&  \times \delta\left(\epsilon^{\prime}-\upsilon_\text{def}\left(\bm{r}^{\prime}\right)\right) n(\bm{r})^{\alpha} \omega\left(\bm{r}, \bm{r}^{\prime}\right) n\left(\bm{r}^{\prime}\right)^{\alpha}    \notag    \\
= & C_{\text{gWT}}^{\text{nloc}} \int d \epsilon d \epsilon^{\prime}\; \widetilde{n}^e(\epsilon)^{\alpha} \widetilde{n}^e\left(\epsilon^{\prime}\right)^{\alpha}  \notag      \\
& \int d \bm{r} d \bm{r}^{\prime} \delta(\epsilon-\upsilon_\text{def}(\bm{r})) \delta\left(\epsilon^{\prime}-\upsilon_\text{def}\left(\bm{r}^{\prime}\right)\right) \omega\left(\bm{r}, \bm{r}^{\prime}\right)       \notag   \\
=& C_{\text{gWT}}^{\text{nloc}} \int d \epsilon  d \epsilon^{\prime}\; \widetilde{n}^e(\epsilon)^{\alpha}  \omega^e \left(\epsilon, \epsilon^{\prime}\right)
\label{eq:E_gWT_e} \widetilde{n}^e\left(\epsilon^{\prime}\right)^{\alpha}
\end{align} }
Note that the integral kernel $\omega(\bm{r},\bm{r}^\prime)$ in the generalized WT functional is represented with the energy coordinate in Eq. (\ref{eq:E_gWT_e}).  It is readily recognized that the transformation from Eq. (\ref{eq:E_gWT}) to Eq. (\ref{eq:E_gWT_e}) can be performed without approximations as long as Eq. (\ref{eq:constancy}) holds exactly. \textcolor{black}{In practice, however, $\widetilde{n}(\epsilon)$ in Eq. (\ref{eq:E_gWT_e}) is taken as the average electron density of the region with coordinate $\epsilon$, thus, 
\begin{equation}  
\widetilde{n}(\epsilon) = \Omega(\epsilon)^{-1} \int d\bm{r}\; n(\bm{r}) \delta(\epsilon - \upsilon_\text{def}(\bm{r})) 
\end{equation}
where $\Omega(\epsilon)$ is the volume of the spatial region with the energy coordinate $\epsilon$ and is given by
\begin{equation} 
\Omega(\epsilon) = \int d\bm{r}\; \delta(\epsilon - \upsilon_\text{def}(\bm{r}))
\end{equation} }
The validity of projecting the kernel $\omega(\bm{r},\bm{r}^\prime)$ and the density $n(\bm{r})$ onto the energy coordinate is guaranteed by the framework of the DFT using the distribution $n^e(\epsilon)$ where the one-to-one correspondence is established\cite{Takahashi2018jpb} between a certain subset of $n^e(\epsilon)$ and a subset of the external potential $\upsilon^e(\epsilon)$ defined with Eq. (\ref{eq:epsilon}). 

It is also worth discussing the numerical advantage of taking the projection of the kernel $\omega$. In a numerical implementation, the LRF for some reference system should be somehow inverted to obtain the kinetic potential. However, it is probably infeasible in practice to make the inversion of the response function $\chi(\bm{r}, \bm{r^\prime})$ represented on the real space since it has six variables at each matrix element. A possible way to overcome the problem is to introduce some basis functions to represent the response function so that one can reduce the size of the matrix. However, the basis functions suitable for the expansion of $\chi$ is not known to the best of our knowledge.        

We close this subsection with a brief summary. Provided that the nonlocal functional model in Eq. (\ref{eq:E_gWT}) is adequate for some applications, the functional projected onto the energy coordinate (Eq. (\ref{eq:E_gWT_e})) will also be applicable since the electron density is reasonably supposed to be nearly constant on the isosurfaces of energy coordinates. In the next subsection, we provide an explicit form of a kinetic energy functional represented on the coordinate $\epsilon$ although it cannot be used for general purposes.    

\subsection{Nonlocal Kinetic Energy in the Energy Representation} 
In this subsection we develop a kinetic energy functional represented on the energy coordinate. We note that the functional will refer to the linear-response function(LRF) of some inhomogeneous electron density.  In the previous works, the nonlocal term $E_\text{WT}^\text{nloc}[n]$ defined in Eq. (\ref{eq:E_WT}) and its sophisticated forms were tested through several applications and their efficiencies were demonstrated to a certain extent.  However, it is not known whether the same form will be as effective a functional as the functional that utilizes the LRF of an inhomogeneous system. We, thus, propose a prototype of the kinetic energy functional with another form although the applicability to general systems might be lost.  

To do this we introduce a parameter $\lambda$ which couples the reference density $n_0(\bm{r})$ and the target density $n_1(\bm{r})$. In the construction of the functional it is presumed that the corresponding kinetic potential $\upsilon_\text{kin}[n_0](\bm{r})$ and the energy $E_\text{kin}[n_0]$ for the given density $n_0$ are somehow provided from the outset. This is the major drawback of the present approach since the preparation of the reasonable $\upsilon_\text{kin}[n_0](\bm{r})$ and $E_\text{kin}[n_0]$ can be a difficult problem by itself. However, there have been a lot of functionals that yield better kinetic potentials and energies for the non-selfconsistent densities. The choice of the functional suitable for our functional is not a subject in the present work and it will be examined in a forthcoming issue. Anyway, the kinetic energy $E_\text{kin}[n_1]$ can be formulated through the coupling parameter integration,    
\begin{align}
E_{\mathrm{kin}}\left[n_{1}\right] &=E_{\mathrm{kin}}\left[n_{0}\right]+\int_{0}^{1} d \lambda \frac{d E_{\mathrm{kin}}\left[n_{\lambda}\right]}{d \lambda}     \notag         \\
& =E_{\mathrm{kin}}\left[n_{0}\right]+\int_{0}^{1} d \lambda \int d \bm{r} \frac{d n_{\lambda}(\bm{r})}{d \lambda} \frac{\delta E_{\mathrm{kin}}\left[n_{\lambda}\right]}{\delta n_{\lambda}(\bm{r})}      \notag           \\
& =E_{\mathrm{kin}}\left[n_{0}\right]+\int d \bm{r} \left( \left(n_{1}(\bm{r})-n_{0}(\bm{r})\right) \right.    \notag      \\
& \;\;\;\;\;\;\;\;\;\;\;\;\;\;\;\;\;\;\;\;\;\;\;\;  \left. \times \int_{0}^{1} d \lambda \frac{\delta E_{\mathrm{kin}}\left[n_{\lambda}\right]}{\delta n_{\lambda}(\bm{r})} \right)
\label{eq:coupling}
\end{align}
In deriving the right hand side of the third equality in Eq. (\ref{eq:coupling}), it is assumed that the density varies linearly with respect to $\lambda$, 
\begin{equation}
n_{\lambda}(\bm{r})=(1-\lambda) n_{0}(\bm{r})+\lambda n_{1}(\bm{r})
\label{eq:linearlity}
\end{equation}
\textcolor{black}{Note that Eq. (\ref{eq:linearlity}) is implicitly based on the premise that there exists the corresponding effective potential $\upsilon_\text{eff}[n_\lambda](\bm{r})$ which yields the intermediate density $n_{\lambda}(\bm{r})$ obeying Eq. (\ref{eq:linearlity}). This is reasonably justified by the one-to-one correspondence\cite{rf:parr_yang_eng} between the potential and the density $n_\lambda$ although the $\upsilon$-representability of  $n_{\lambda}(\bm{r})$ is not guaranteed.} We further approximate that the kinetic potential $\upsilon_{\text{kin}}\left[n_{\lambda}\right](\bm{r})=\frac{\delta E_{\mathrm{kin}}\left[n_{\lambda}\right]}{\delta n_{\lambda}(\bm{r})}$ in Eq. (\ref{eq:coupling}) changes linearly. Then, its integration with respect to $\lambda$ becomes
\begin{equation}
\int_{0}^{1} d \lambda \upsilon_{\mathrm{kin}}\left[n_{\lambda}\right](\bm{r})=\frac{1}{2}\left(\upsilon_{\mathrm{kin}}\left[n_{0}\right](\bm{r})+ \upsilon_{\mathrm{kin}}\left[n_{1}\right](\bm{r})\right)
\label{eq:v_kin_coupling}
\end{equation}
After the substitution of Eq. (\ref{eq:v_kin_coupling}) into Eq. (\ref{eq:coupling}), we take the functional derivatives with respect to $n_1(\bm{r})$ for both sides of Eq. (\ref{eq:coupling}), which reads    
\begin{align}
\upsilon_{\mathrm{kin}}\left[n_{1}\right](\bm{r}) &= \upsilon_{\mathrm{kin}}\left[n_{0}\right](\bm{r})      \notag     \\
&+ \int d \bm{r}^{\prime} \frac{\delta \upsilon_{\mathrm{kin}}\left[n_{1}\right](\bm{r})}{\delta n_{1}\left(\bm{r}^{\prime}\right)}\left(n_{1}\left(\bm{r}^{\prime}\right)-n_{0}\left(\bm{r}^{\prime}\right)\right)     
\label{eq:v_kin}
\end{align}
In Eq. (\ref{eq:v_kin}) we find the functional derivative of the kinetic potential $\upsilon_\text{kin}[n_1](\bm{r})$, which can be replaced by the negative inverse of the response function as shown in Eq. (\ref{eq:vkin_der}). Therefore, the kinetic potential can be rewritten as 
\begin{align}  
\upsilon_{\mathrm{kin}}\left[n_{1}\right](\bm{r}) &= \upsilon_{\mathrm{kin}}\left[n_{0}\right](\bm{r})       \notag    \\
&- \int d \bm{r}^{\prime} \chi_{0}^{-1}\left(\bm{r}, \bm{r}^{\prime}\right)\left(n_{1}\left(\bm{r}^{\prime}\right)-n_{0}\left(\bm{r}^{\prime}\right)\right)
\label{eq:v_kin2}
\end{align}
where $\chi_0$ is the LRF for a reference system with the density $n_0(\bm{r})$ and no longer supposed to be that for the HEG. It is readily proved that the kinetic energy $E_\text{kin}[n_1]$ is obtained as 
\begin{align}
E_\text{kin}\left[n_{1}\right] &= E_\text{kin}\left[n_{0}\right] +\int d \bm{r} \upsilon_\text{kin}\left[n_{0}\right](\bm{r}) \delta n(\bm{r})            \notag       \\
& \;\;\;\;\;\;\;\;\;  -\frac{1}{2} \int d \bm{r} d \bm{r}^{\prime} \chi_{0}^{-1}\left(\bm{r}, \bm{r}^{\prime}\right) \delta n(\bm{r}) \delta n\left(\bm{r}^{\prime}\right)
\label{eq:Ekin}
\end{align}
where  $\delta n(\bm{r})$ is defined as
\begin{equation}
\delta n(\bm{r}) = n_1(\bm{r}) - n_0(\bm{r})
\end{equation}
We note that Eq. (\ref{eq:Ekin}) has the desirable property that its second derivative with respect to the density $n(\bm{r})$ becomes the negative inverse of the LRF for some reference system. 
As a major issue in the present development, we represent the nonlocal term in Eq. (\ref{eq:v_kin2}) on the energy coordinate, 
\begin{align}
\upsilon_{\mathrm{kin}}^{e}\left[n_{1}\right](\bm{r}) &=\upsilon_{\mathrm{kin}}\left[n_{0}\right](\bm{r})       \notag      \\
&- \int d \epsilon^{\prime}  \chi_{0}^{e}\left(\epsilon, \epsilon^{\prime}\right)^{-1}\left(n_{1}^{e}\left(\epsilon^{\prime}\right)-n_{0}^{e}\left(\epsilon^{\prime}\right)\right)
\label{eq:v_kin_e}
\end{align}
$\chi_{0}^{e}\left(\epsilon, \epsilon^{\prime}\right)$ in Eq. (\ref{eq:v_kin_e}) can be obtained by projecting the LRF in the real space onto the energy coordinate,  
\begin{equation} 
\chi_{0}^{e}\left(\epsilon, \epsilon^{\prime}\right) = \int d \bm{r} d \bm{r}^{\prime} \chi_{0}\left(\bm{r}, \bm{r}^{\prime}\right) \delta(\epsilon-v(\bm{r})) \delta\left(\epsilon^{\prime}-v\left(\bm{r}^{\prime}\right)\right)
\label{eq:chi_e}
\end{equation}
Note that $\chi_{0}^{e}\left(\epsilon, \epsilon^{\prime}\right)^{-1}$ represents an element of the inverse matrix of $\chi_{0}^{e}$. Accordingly, the kinetic energy functional $E_\text{kin}\left[n\right]$ in Eq. (\ref{eq:Ekin}) is transformed to  
\begin{align}
E_\text{kin}^e \left[n_{1}\right] &= E_\text{kin}\left[n_{0}\right] +\int d \bm{r} \upsilon_\text{kin}\left[n_{0}\right](\bm{r}) \delta n(\bm{r})            \notag       \\
& \;\;\;\;\;\;\;\;\;  -\frac{1}{2} \int d \epsilon d \epsilon^{\prime} \chi_{0}^{e}\left(\epsilon, \epsilon^{\prime}\right)^{-1} \delta n^e(\epsilon) \delta n^e\left(\epsilon^{\prime}\right)
\label{eq:Ekin_e}
\end{align}
As described in Appendix A, it is possible to prove that $\chi_{0}^{e}\left(\epsilon, \epsilon^{\prime}\right)$ is positive semi-definite and hence, invertible through the Moore-Penrose pseudo-inverse method. The detailed discussion on the eigenvector with the null eigenvalue of the matrix is shown in Ref. \onlinecite{Matubayasi2003jcp}. Thus, only the nonlocal term in the kinetic potential is described with the energy representation.  In Eq. (\ref{eq:v_kin_e}) it should be noticed that the first term in the right hand side is the function of $\bm{r}$, while the second term depends on the energy coordinate $\epsilon$. It doesn't matter, however, because the kinetic potential on a spatial coordinate $\bm{r}$ can be readily obtained through the transformation $\epsilon = \upsilon_\text{def}(\bm{r})$ using Eq. (\ref{eq:epsilon}).

\subsection{Self-consistent Field Calculation for Orbital-free Density-functional Theory} 
There have been a lot of works devoted to expedite the self-consistent field (SCF) calculations in OF-DFT. However, in the present development we are not interested in the implementation of efficient SCF techniques. Here, we devise a minimal procedure to achieve SCF convergence in our approach. A possible improvement of the algorithm will be the subject of a forthcoming issue. We start the discussion with the variational principle for the energy functional $E[n]$. As shown in Eq. (\ref{eq:E_der1}), the minimization of the energy $E[n]$ with respect to $n(\bm{r})$ under the constraint that the total number of electrons is fixed at $N$ gives the chemical potential $\mu$ for the electrons at the stationary state. That is, the potential $\upsilon[n](\bm{r})$ defined by $\upsilon[n](\bm{r}) = \frac{\delta E[n]}{\delta  n(\bm{r})}$ is constant irrespective of $\bm{r}$ at the energy minimum. Provided that the chemical potential $\mu$ is set at 0, minimization of the functional $E[n]$ is equivalent to the solution of the nonlinear equation, 
\begin{equation}   
\upsilon[n](\bm{r}) = 0
\end{equation}
Within the first-order variation of $n$, the solution $\overline{n}$ of the equation satisfies the following approximation	
\begin{align}
\upsilon\left[\overline{n}\right](\bm{r}) & \cong \upsilon[n](\bm{r})      \notag   \\
& +\int d \bm{r}^{\prime} \frac{\delta \upsilon[n](\bm{r})}{\delta n\left(\bm{r}^{\prime}\right)}\left(\overline{n}\left(\bm{r}^{\prime}\right)-n\left(\bm{r}^{\prime}\right)\right)=0
\end{align}
where $n$ is an arbitrary density reasonably close to $\overline{n}$. Then, the formal solution for the above equation is given by
\begin{equation}
\overline{n}(\bm{r}) \cong n(\bm{r})-\int d \bm{r}^{\prime} \frac{\delta n(\bm{r})}{\delta \upsilon[n]\left(\bm{r}^{\prime}\right)} \upsilon[n]\left(\bm{r}^{\prime}\right)
\label{eq:ovl_n}
\end{equation}
Thus, the electron density $n_i$ at the $i$th SCF step can be updated as
\begin{equation}
n_{i+1}(\bm{r}) = n_{i}(\bm{r}) -\int d \bm{r}^{\prime} \frac{\delta n_i(\bm{r})}{\delta \upsilon[n_i]\left(\bm{r}^{\prime}\right)} \upsilon[n_i]\left(\bm{r}^{\prime}\right)
\label{eq:SCF}  
\end{equation}
Unfortunately, the SCF iteration through Eq. (\ref{eq:SCF}) is practically infeasible because the functional derivative of the density with respect to the potential $\upsilon[n](\bm{r})(=\upsilon_\text{kin}+\upsilon_\text{eff})$ cannot be readily obtained. A simple but inefficient way to bypass the calculation of the integral kernel in Eq. (\ref{eq:SCF}) is to replace the kernel with a unit matrix multiplied by a small positive factor $\eta$. In our preliminary calculations, however, it was found that such an approach gives rise to quite slow convergence in the density. In the present implementation, we replace the kernel with the derivative of the density with respect to the kinetic potential, thus, 
\begin{equation}  
n_{i+1}(\bm{r}) = n_{i}(\bm{r}) - \eta \int d \bm{r}^{\prime} \frac{\delta n_i(\bm{r})}{\delta \upsilon_\text{kin}[n_i]\left(\bm{r}^{\prime}\right)} \upsilon[n_i]\left(\bm{r}^{\prime}\right)
\label{eq:SCF2}
\end{equation}
It is important to note that the replacement of the kernel doesn't change the stationary condition. Actually, at the convergence of the density, that is, $n_{i+1}(\bm{r}) - n_{i}(\bm{r})  \cong 0$, we have, 
\begin{equation}
\int d \bm{r}^{\prime} \frac{\delta n_i(\bm{r})}{\delta \upsilon_\text{kin}[n_i]\left(\bm{r}^{\prime}\right)} \upsilon[n_i]\left(\bm{r}^{\prime}\right) \cong 0
\end{equation}
Then, $\upsilon[n_i]\left(\bm{r}^{\prime}\right) \cong 0$ is guaranteed since the kernel $\frac{\delta n_i(\bm{r})}{\delta \upsilon_\text{kin}[n_i]\left(\bm{r}^{\prime}\right)}$ is always positive semidefinite and invertible.  Further, we assume that the kernel in Eq. (\ref{eq:SCF2}) is evaluated for the reference system and represented on the energy coordinate, 
\begin{equation}
n_{i+1}(\bm{r})=n_{i}(\bm{r}) + \eta \int d \epsilon^{\prime} \chi_{0}^{e}\left(\epsilon, \epsilon^{\prime}\right) \upsilon\left[n_{i}\right]\left(\epsilon^{\prime}\right)
\label{eq:SCF_erep}
\end{equation}
where the potential $\upsilon[n](\epsilon)$ is defined as
\begin{equation}
\upsilon[n](\epsilon) = \int d\bm{r} \delta (\epsilon - \upsilon_\text{def}(\bm{r}))\upsilon[n](\bm{r})
\end{equation}
In Eq. (\ref{eq:SCF_erep}) it is possible that the kernel $\chi_{0}^{e}\left(\epsilon, \epsilon^{\prime}\right)$ plays as a nonlocal weight function acting on the potential to expedite the SCF. \textcolor{black}{The choice of the value $\eta$ will affect more or less the rate of the SCF convergence and  accuracy of the results. In the present work, the appropriate value of $\eta$ is determined empirically as $\eta = 0.05$, and it is used in the SCF calculations throughout.}  

Finally, we describe a minor issue for the density in the SCF iteration. It is common to employ the variable $\varphi(\bm{r})$, that is, the square root of the density $n(\bm{r})$,   
\begin{equation}
\varphi(\bm{r}) = n(\bm{r})^{\frac{1}{2}} 
\end{equation}
to avoid the situation that $n(\bm{r})$ becomes negative during the SCF. Then, the SCF iteration using Eq. (\ref{eq:SCF2}) should be modified as 
\begin{align}
\varphi_{i+1}(\bm{r}) & =\varphi_{i}(\bm{r})       \notag      \\
& \;\;\;\;\;\; - \eta \int d \bm{r}^{\prime}\left\{\frac{\delta \varphi(\bm{r})}{\delta n(\bm{r})} \frac{\delta n(\bm{r})}{\delta \upsilon_{\mathrm{kin}}[n]\left(\bm{r}^{\prime}\right)}\right\}_{n_{i}} \upsilon[n_i]\left(\bm{r}^{\prime}\right)
\end{align}
Accordingly, Eq. (\ref{eq:SCF_erep}) is rewritten as 
\begin{equation}
\varphi_{i+1}(\bm{r})=\varphi_{i}(\bm{r})+ \eta \int d \epsilon^{\prime} \widetilde{\chi}_{0}^{e}\left(\epsilon, \epsilon^{\prime}\right) \upsilon\left[n_{i}\right]\left(\epsilon^{\prime}\right)
\label{eq:SCF_e}
\end{equation}
where $\widetilde{\chi}_{0}^{e}$ is defined as
\begin{align}
\widetilde{\chi}_{0}^{e}\left(\epsilon, \epsilon^{\prime}\right) & = \int d \bm{r} d \bm{r}^{\prime}\left\{\frac{\delta \varphi(\bm{r})}{\delta n(\bm{r})} \frac{\delta n(\bm{r})}{\delta \upsilon_{\mathrm{kin}}\left(\bm{r}^{\prime}\right)}\right\}_{n_{0}}       \notag     \\
& \;\;\;   \times  \delta\left(\epsilon-\upsilon_{\mathrm{def}}(\bm{r})\right) \delta\left(\epsilon^\prime-\upsilon_{\mathrm{def}}\left(\bm{r}^{\prime}\right)\right)
\label{eq:tilde_chi}
\end{align}
In Eq. (\ref{eq:tilde_chi}), $n_0$ is the density of the reference system.  

\section{Computational Details}
As a benchmark test we apply the kinetic energy functional of Eq. (\ref{eq:Ekin_e}) to pseudo atomic systems(H, He, Ne, and Ar), where the core charge $Z_\upsilon$ of each atom is shifted by $q$ within the range of $ -0.2 \leq q \leq  +0.2$ in the unit of the elementary charge. Throughout the calculations, the system with $q=0$ is taken as the reference system to build the response function $\chi_0^e(\epsilon, \epsilon^\prime)$ as well as the density $n_0^e(\epsilon)$ on the energy coordinate.  Then, the variation of the kinetic energy of each atom is compared with that given by the Kohn-Sham DFT calculation. We also employ other OF-DFT functionals, that is, $E_\text{TF}[n]$ in Eq. (\ref{eq:E_TF}) and $E_\text{TFW}[n]$ in Eq. (\ref{eq:TF+lvW}) for comparisons. In these calculations the electron density $n(\bm{r})$ is individually optimized beforehand in the separate Kohn-Sham DFT calculation. 

Next, we perform the variational calculation using the kinetic potential $v_{\text{kin}}^{e}\left[n\right](\bm{r})$ in Eq. (\ref{eq:v_kin_e}) to obtain the self-consistent electron density, where the reference density $n_0(\bm{r})$ is employed as an initial guess. Then, we compute the radial distribution functions $4\pi r^2 n(r)$ of the Ne and Ar systems with $q = 0.2$ through OF-DFT SCF calculations using Eq. (\ref{eq:SCF_e}).  

The first subsection provides the details of the base program to perform numerical OF-DFT calculations using the real-space grids. In the second subsection, we introduce the construction of the pseudo atoms that mimic H, He, Ne, and Ar, where the bare Coulomb potentials are replaced by some local pseudopotentials to realize the smooth behaviors of the densities near the atomic cores although the properties of the corresponding atoms are no longer maintained. The numerical details for the construction of the response function and the energy electron density on the energy coordinate is described in the third subsection.

\subsection{Real-space Grid Approach}
The program module for the present OF-DFT is newly developed and embedded in the Kohn-Sham DFT program \textquoteleft Vmol\textquoteright \cite{rf:takahashi2000cl, takahashi2001jpca, rf:takahashi2001jcc, rf:takahashi2004jcp, Takahashi2019jpcb, Takahashi2020jcim}\; based on the real-space grid approach\cite{rf:chelikowsky1994prb,rf:chelikowsky1994prl}. Thus, the electron density $n(\bm{r})$ is represented on the real-space grids uniformly placed within a cubic box. Throughout the present calculations, the size of the real-space cell is set at $L = 18.354$ a.u. and each axis is discretized by equally spaced 64 grids, which leads the grid width $h = 0.2868$ a.u. The exchange-correlation energy $E_{xc}[n]$ in Eq. (\ref{eq:E_decomp}) is evaluated with the BLYP functional\cite{rf:becke1988pra,rf:lee1988prb}.  To make comparisons among the kinetic energy functionals, we also employ the functionals $E_\text{TF}[n]$ and $E_\text{vW}[n]$ given in Eqs. (\ref{eq:E_TF}) and (\ref{eq:E_vW}), respectively. The Laplacian in Eq. (\ref{eq:E_vW}) is represented with the fourth-order finite difference method\cite{rf:chelikowsky1994prb,rf:chelikowsky1994prl}. 

\subsection{Construction of Pseudo Atoms}
\begin{figure}[h]
\centering
\scalebox{0.35}[0.35] {\includegraphics[trim=10 10 10 10,clip]{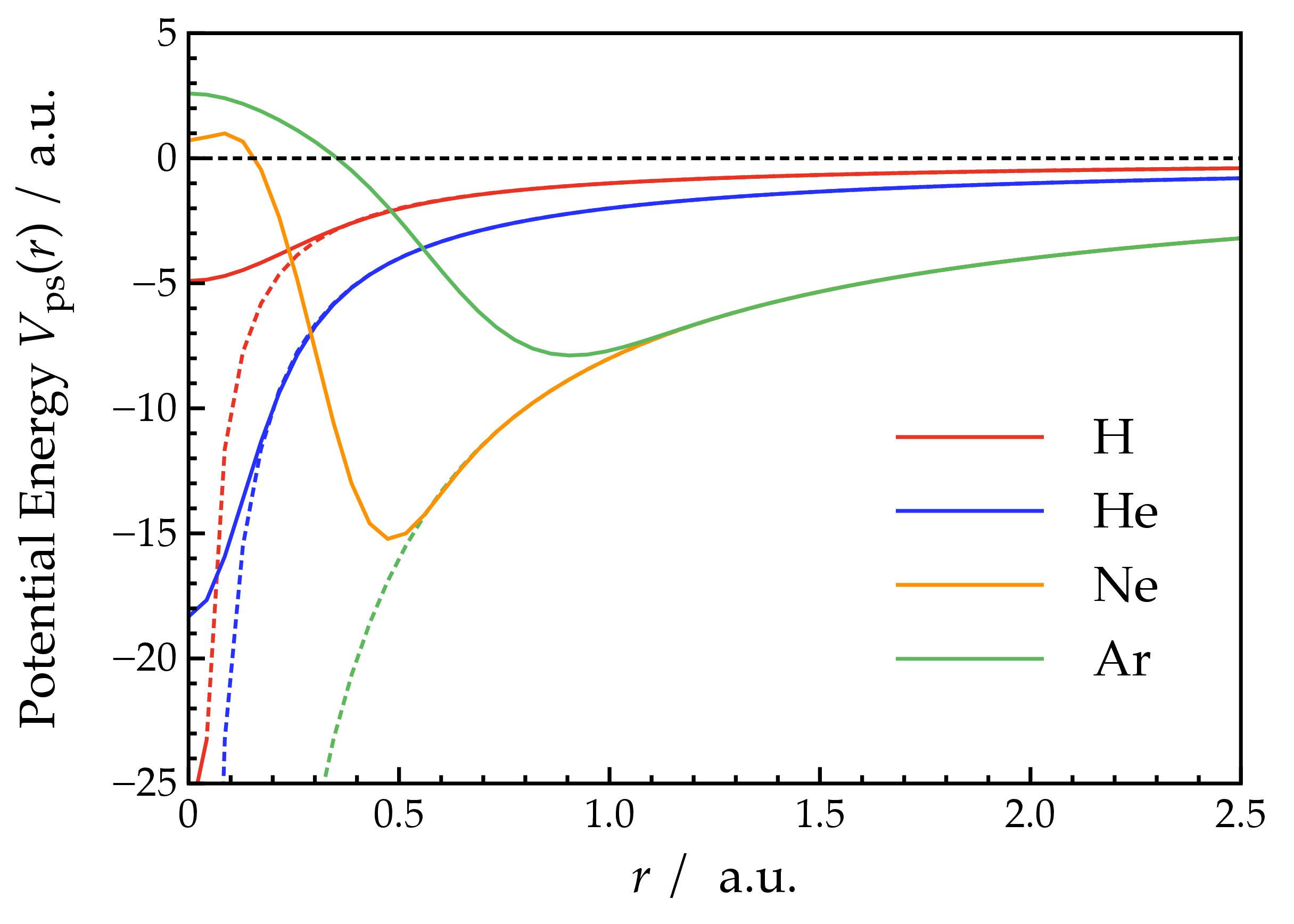}}            
\caption{\label{VPS}  The local pseudopotentials constructed using Eq. (\ref{eq:BHS}) for H, He, Ne, and Ar pseudo atoms. The broken lines are the corresponding Coulomb potentials of the valence charges $Z_{v}$. }
\end{figure}

To achieve the OF-DFT calculation within the plane-wave or real-space grid approach, it is also a subject of critical importance to develop \textit{local} pseudopotentials for atoms. Unfortunately, to the best of our knowledge, there have not been sufficient amount of works along this line. In the present study, however, we are not interested in the applications to realistic atoms. We, thus, consider \textquoteleft pseudo\textquoteright\; atoms which stem from Hydrogen, Helium, Neon, and Argon atoms. The sole purpose of the pseudization is to realize the smooth variation of the density so that the real-space grid approach can adequately describe it. To this end we merely adopt the local term $V_\text{loc}(r)$ in an existing pseudopotential set as a function of the distance $r$ between the electron and the nucleus. As a specific treatment for pseudo Ne and Ar atoms, we additionally embed the potential $\Delta V_\text{s}(r)$ assigned as the S component in the nonlocal term to mimic the repulsive potential due to the core electrons. Actually, we found that the valence electrons in Ne and Ar erroneously penetrate into the core region without the term $\Delta V_\text{s}(r)$. In the present calculation, we use the terms in the BHS pseudopotential\cite{rf:bachelet1982prb} as components to build the potential. We summarize the form of the local pseudopotential $V_\text{ps}^\text{A}(r)$ for atom A(= H, He, Ne, or Ar) used in our calculations,    
\begin{equation}           
\left\{\begin{array}{l}
V_\text{ps}^\text{A}(r)=V_\text{loc}^\text{A}(r) \quad \text{(A = H, He)} \\
\\
V_\text{ps}^\text{A}(r)=V_\text{loc}^\text{A}(r) + \Delta V_\text{s}^\text{A}(r) \quad \text{(A = Ne, Ar)}
\end{array}\right.
\label{eq:BHS}
\end{equation}
where $V_\text{loc}^\text{A}(r)$ and $\Delta V_{s}^\text{A}(r)$ are, respectively, given in the forms,
\begin{equation}
V_\text{loc}(r) = -\frac{Z_{v}}{r}\left[\sum_{i=1}^{2} c_{i} \operatorname{erf}\left[\left(\alpha_{i}\right)^{1 / 2} r\right]\right]
\label{eq:vloc_BHS}
\end{equation}
and
\begin{equation}
\Delta V_\text{s}(r) = \sum_{i=1}^{3}\left(P_{i} + r^{2} P_{i+3}\right) \exp(-\alpha_{i} r^{2})
\label{eq:vnloc_BHS}
\end{equation}
In Eqs. (\ref{eq:vloc_BHS}) and (\ref{eq:vnloc_BHS}) we omit the superscript A for simplicity. In Eq. (\ref{eq:vloc_BHS}) $Z_{v}$ is the corresponding valence charge for atom A. $Z_{v}$ are 1.0 and 2.0 for H and He atoms, respectively, while $Z_{v} = 8.0$ for Ne and Ar. The coefficients $c_i$ and $\alpha_i$ in Eq. (\ref{eq:vloc_BHS}) are tabulated in Ref. \onlinecite{rf:bachelet1982prb} for various atoms. The two sets of coefficients $P_i$ and $P_{i+3}$ in Eq. (\ref{eq:vnloc_BHS}) correspond to the exponents $\alpha_i\; (i=1,2,3)$. The set of $P_i$ can be obtained by the numerical reconstruction of the data in Table IV in Ref. \onlinecite{rf:bachelet1982prb}. We refer the readers to the paper for the detail of the algorithm. The local pseudopotentials employed in the present calculations are displayed in Fig. \ref{VPS}. In the graphs for Ne and Ar atoms, it is shown that each $V_\text{ps}(r)$ rises notably near the core region due to the repulsive potential $\Delta V_\text{s}(r)$.  
It should be stressed that the pseudopotentials given in Eq. (\ref{eq:BHS}) are \textit{not} optimized to reproduce the properties of the corresponding realistic atoms. Hence, the eigenvalues provided by the Kohn-Sham DFT calculation are not coincident with those constructed by the full set of pseudopotential that includes the nonlocal terms. However, it makes complete sense to compare the result given by OF-DFT with that given by the KS-DFT calculation.  The time-saving double-grid technique\cite{rf:ono1999prl} is employed to describe adequately the steep variation of the local pseudopotentials in the core region, where the width of the dense grid is set at $h/5$ throughout the calculations.     

\subsection{Energy Coordinate and Linear-response Function}
As shown in Eq. (\ref{eq:epsilon}) the bare Coulomb potential is usually employed as the defining potential $\upsilon_\text{def}(\bm{r})$ for the energy coordinate $\epsilon$. However, another potential $\upsilon_\text{def}^\prime(\bm{r})$ can also be adopted as a defining potential as long as the relation $\upsilon_\text{def}^\prime(\bm{r}_\text{A}) = \upsilon_\text{def}^\prime(\bm{r}_\text{B})$ holds for any points $\bm{r}_\text{A}$ and $\bm{r}_\text{B}$ satisfying $\upsilon_\text{def}(\bm{r}_\text{A}) = \upsilon_\text{def}(\bm{r}_\text{B})$\cite{Takahashi2018jpb}. In the present development, in contrast to the previous works\cite{Takahashi2018jpb,Takahashi2020}, we employ the opposite sign of the BHS local potential $V_\text{loc}^\text{BHS}(r)$ as the defining potential, i.e., $\upsilon_\text{def}(\bm{r}) = -V_\text{loc}^\text{BHS}(r)$. 

The log-scaled energy coordinate with the range of $(\log \epsilon_\text{min}, \log \epsilon_\text{max})$ is uniformly discretized by $N^e_\text{grid}$ grid points. Then, the energy electron density $n^e(\epsilon)$ in Eq. (\ref{eq:n_e}) and the response function $\chi_0^e(\epsilon, \epsilon^\prime)$ in Eq. (\ref{eq:chi_e}) are numerically constructed on the discrete energy coordinates. The parameters $N_\text{grid}^e$ and $(\epsilon_\text{min}, \epsilon_\text{max})$ are individually determined for each atom and the values are summarized in the \textquoteleft Supplementary Material\textquoteright. 

The linear-response function(LRF) $\chi_0(\bm{r}, \bm{r}^\prime)$ in Eq. (\ref{eq:chi_e}) for a reference system is obtained by the 2nd-order perturbation theory. To do this, we first solve a Kohn-Sham equation for the reference system(i.e., $q = 0.0$),    
\begin{equation}
\left(-\frac{1}{2} \nabla^{2}+\upsilon_\text{eff}\left[n_{0}\right](\bm{r}) \right) \varphi_{i}^{0}(\bm{r})=\epsilon_{i}^{0} \varphi_{i}^{0}(\bm{r})
\label{eq:KS_eq}
\end{equation}
where $n_0$ is the electron density constructed from the self-consistent solutions $\{\varphi_i^0\}$, and $\{\epsilon_i^0\}$ is the corresponding eigenvalues. 
Then, the LRF is evaluated as
\begin{align}
\chi_0(\bm{r},\bm{r}^\prime) &= \sum_{i}^{\text {occ}} \sum_{a}^{\text {vir}} \frac{1}{\varepsilon_{a}^{0}-\varepsilon_{i}^{0}}\varphi_{i}^{0*}(\bm{r}) \varphi_{a}^{0}(\bm{r})      \notag           \\
& \;\;\;\;\;\;\;\;  \times \varphi_{a}^{0*}(\bm{r}^\prime) \varphi_{i}^{0}(\bm{r}^\prime)
\label{eq:PT2}
\end{align}
where the indices $i$ and $a$ are for the occupied and virtual orbitals, respectively. \textcolor{black}{Explicitly, the numbers of the virtual orbitals included in Eq. (\ref{eq:PT2}) are 23 for H and He, 20 for Ne, and 22 for Ar. It is found that the LRF is rather insensitive to the number of the virtual orbitals. However, attention should be paid to the upper limit since the virtual orbitals with larger eigenvalues can not be adequately accommodated in the real-space cell. } Note that the kinetic potential $\upsilon_\text{kin}[n_0](\bm{r})$ in Eq. (\ref{eq:v_kin_e}) as well as $E_\text{kin}[n_0]$ in Eq. (\ref{eq:Ekin_e}) for the reference system can also be obtained from the solution of Eq. (\ref{eq:KS_eq}). Specifically for $\upsilon_\text{kin}[n_0](\bm{r})$, we use the relation,  
\begin{equation}
\upsilon_\text{kin}[n_0](\bm{r}) = - \upsilon_\text{eff}\left[n_{0}\right](\bm{r})
\end{equation}  
Although the response function $\chi_0(\bm{r},\bm{r}^\prime)$ can be easily computed, the projection onto the energy coordinate through Eq. (\ref{eq:chi_e}) is rather time-consuming. This could be a serious problem when we consider its applications to larger systems in a later development. In our perspective, however, the LRF of a composite system might be adequately constructed from the overlap of the LRF of the constituent systems on the energy coordinate.  

\subsection{Inverse of Linear-response Function}
We also make a remark on the inversion of the LRF defined on the energy coordinate. As shown in Appendix A, the matrix $\chi_0^e(\epsilon,\epsilon^\prime)$ is positive semi-definite, and hence can be inverted using the pseudo-inverse method. Here, we discuss another issue associated with the inversion. Suppose that the matrix is given in the form of the spectral decomposition,
\begin{equation}
\chi_0^e(\epsilon,\epsilon^\prime) = \sum_i^{N^e_\text{grid}} \left\langle\epsilon \mid \gamma_{i}\right\rangle g_{i}\left\langle\gamma_{i} \mid \epsilon^{\prime}\right\rangle
\label{eq:LRF_SD}
\end{equation}
where $\gamma_{i}(\epsilon)$ is the eigenvector of the matrix $\chi_0^e$, and $g_i$ is the corresponding eigenvalue. The matrix includes at least one null eigenvector with the 0 eigenvalue due to the condition that the number of electrons is being fixed. However, it is found unexpectedly that there exists a lot of eigenvectors with substantially zero eigenvalues for the atomic systems studied in the present work. In addition, it is also revealed that the eigenvector with the largest eigenvalue dominates the construction of the matrix. 
\begin{figure}[h]
\centering
\scalebox{0.44}[0.44] {\includegraphics[trim=0 95 10 10,clip, angle=-90]{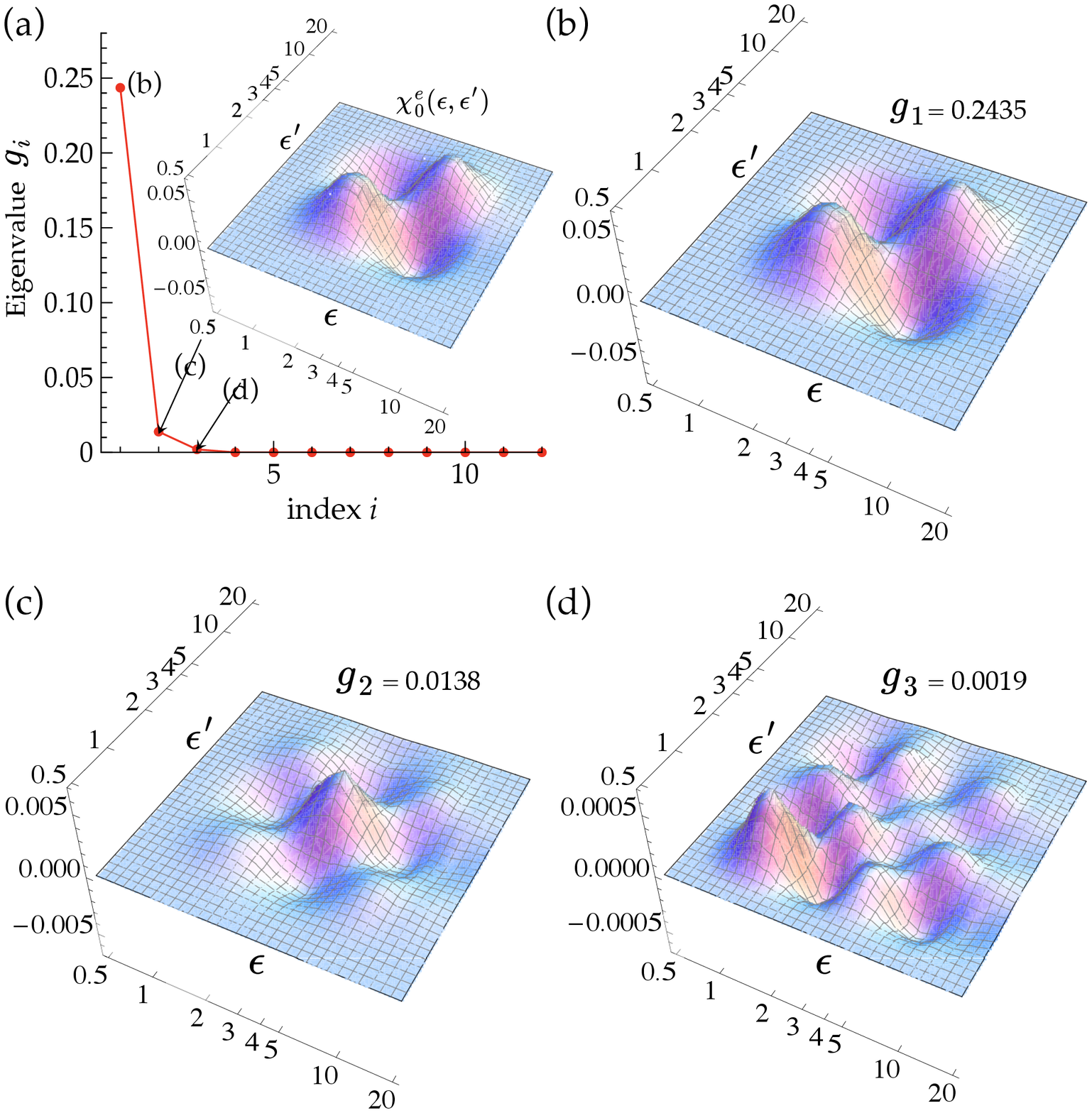}}            
\caption{\label{LRF}  (a) The plot of the eigenvalues $\{g_i\}$ in descending order for the pseudo Ar atom. The linear-response function $\chi_0^e(\epsilon,\epsilon^\prime)$ of the Ar atom is superposed. (b), (c), and (d) show the functions $g_i \gamma_i(\epsilon)\gamma_i(\epsilon^\prime)$ (see Eq. (\ref{eq:LRF_SD})) for the indices $i=1$, 2, and 3, respectively. Units of the energy coordinate $\epsilon$ is a.u., while that of $\chi_0^e(\epsilon,\epsilon^\prime)$ is a.u.$^{-2}$.  }
\end{figure} 
Shown in Fig. \ref{LRF}(a) is the eigenvalues $\{g_i\}$ plotted for the pseudo Ar atom in descending order, where the three-dimensional graph for the LRF $\chi_0^e(\epsilon,\epsilon^\prime)$ is also superposed. The notations (b), (c), and (d) in the plot denote the three largest eigenvalues, for which the corresponding LRF components $g_i\gamma_i(\epsilon)\gamma_i(\epsilon^\prime)$ are drawn in Figs. \ref{LRF}(b), \ref{LRF}(c) and \ref{LRF}(d), respectively. By consulting Figs. \ref{LRF}(a) and \ref{LRF}(b), it is recognized that the eigenvector $\gamma_1$ with the lowest frequency is dominant in the decomposition of the total LRF in Eq. (\ref{eq:LRF_SD}). It seems that higher frequency mode $\gamma_i$ gives much smaller contribution to the LRF as observed in the sequence of Figs. \ref{LRF}(b), \ref{LRF}(c) and \ref{LRF}(d).  It is remarkable that the response function generated by the lowest frequency mode $\gamma_1$ faithfully realizes the total LRF shown in Fig. \ref{LRF}(a). It is found that this trend is common to all the atoms treated in the present work. In the pseudo-inverse approach, the inverse of the LRF of Eq. (\ref{eq:LRF_SD}) can be obtained by 
\begin{equation}
\chi_0^{e}(\epsilon,\epsilon^\prime)^{-1} = \sum_i^{N^e_\text{grid}-1} \left\langle\epsilon \mid \gamma_{i}\right\rangle g_{i}^{-1}\left\langle\gamma_{i} \mid \epsilon^{\prime}\right\rangle 
\label{eq:LRF_inv}
\end{equation}
Note that the eigenvector with 0 eigenvalue is to be excluded from the summation in the right hand side of Eq. (\ref{eq:LRF_inv}). 
In a practical calculation, however, the smallest eigenvalue may not become exactly zero due to the numerical error. Thus, the eigenvector with the smallest eigenvalue should be excluded from the construction of the inverse matrix.  
However, it is found in the preliminary calculations that the faithful execution of Eq. (\ref{eq:LRF_inv}) gives rise to the unphysical oscillations in the matrix $\chi_0^{e}(\epsilon,\epsilon^\prime)^{-1}$ due to the high frequency modes multiplied by the huge values $g_i^{-1}$. \textcolor{black}{Actually, as demonstrated in Appendix B, the magnitude of the component of the response function $\chi_0^e(\epsilon, \epsilon^\prime)$ with a small eigenvalue is found to be within the grid error due to the use of the real-space grids in representing the function $\chi_0(\bm{r}, \bm{r}^\prime)$.} Thus, the summation in Eq. (\ref{eq:LRF_inv}) should be truncated after the first few leading terms in practice. Throughout this work, we only consider the eigenvector $\gamma_1$ with the largest eigenvalue $g_1$ to construct the inverse of the LRF. The effect of the inclusion of the residual contributions will be examined later.   

\section{Results and Discussion}
First, we apply our kinetic energy functional $E_\text{kin}^e[n]$ to the pseudo atoms (H, He, Ne, and Ar) to compute the variation of the kinetic energy for each atom when the valence charge $Z_v$ is shifted by $q$ within the range of $-0.2 \leq q \leq 0.2$ in the unit of the elementary charge. The density $n(\bm{r})$ optimized by Kohn-Sham DFT (KS-DFT) is adopted as the argument of the functional.   

Next, the self-consistent densities for the Ne and Ar atoms are computed through the present OF-DFT approach. The radial distributions of the electrons are compared with those obtained by the KS-DFT calculations. 

\subsection{Kinetic Energy Calculation Using Non-self-consistent Density}
\begin{figure}[h]
\centering
\scalebox{0.35}[0.35] {\includegraphics[trim=0 5 10 10,clip, ]{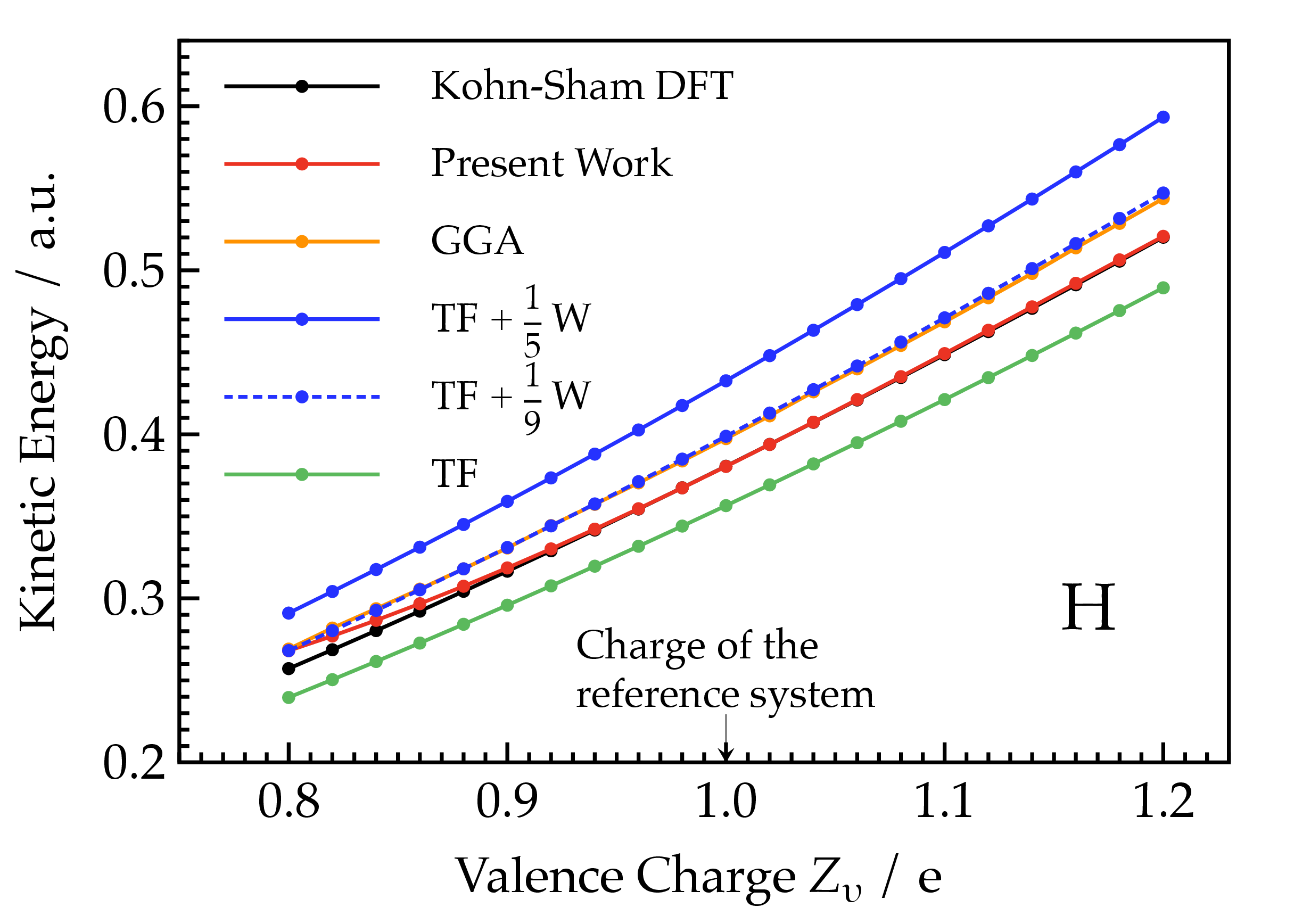}}            
\caption{\label{Pseudo_H}  Plots of the kinetic energies computed by various kinetic functionals for the variation of the valence charge $Z_v$ of the pseudo H atom. The atom with $Z_v = 1.0$ is taken as the reference system. The value of the \textquoteleft Present Work\textquoteright\; is provided using Eq. (\ref{eq:Ekin_e}). The GGA functional is that given in Eq. (\ref{eq:GGA}) or equivalently Eq. (12) in Ref. \onlinecite{Thakkar1992}. \textquoteleft TF + 1/5W\textquoteright\; and \textquoteleft TF + 1/9W\textquoteright\; are obtained by Eq. (\ref{eq:TF+lvW}) with $\lambda = 1/5$ and $1/9$, respectively.  \textquoteleft TF\textquoteright\; is evaluated by Eq. (\ref{eq:E_TF}).  }
\end{figure}

Figure \ref{Pseudo_H} shows the profile of the kinetic energy $E_\text{kin}^e[n]$ (Eq. (\ref{eq:Ekin_e})) with respect to the variation of the valence charge $Z_v\; (0.8 \leq Z_v \leq 1.2)$ of the pseudo H atom. To make comparisons we also employ the Thomas-Fermi (TF) functional $E_\text{TF}$ in Eq. (\ref{eq:E_TF}) and $E_\text{TF}$ combined with the Weizs\"{a}cker correction, that is, the TFW functional $E_\text{TFW}$ in Eq. (\ref{eq:TF+lvW}). The electron density $n(\bm{r})$ adopted as the argument is that optimized through the self-consistent field (SCF) calculation of the KS-DFT method. In every plot, the kinetic energy increases monotonically with respect to the increase in the valence charge $Z_v$. The increase in $Z_v$ causes the stabilization of the potential energy of the system, which leads to the increase in the kinetic energy. As observed in the figure, the functional $E_\text{kin}^e[n]$ developed in the present work shows excellent agreement with the result of KS-DFT except for the region around $ Z_v = 0.80$. Although the kinetic energy at $ Z_v =\;0.80$ varies by $\sim0.13$ a.u. from that of the reference system with $ Z_v =\;1.0$, the deviation of the present work from the KS-DFT value stays $\sim 0.01$ a.u. It is notable that  the present work perfectly realizes the KS-DFT values around another end of the valence charge, i.e., $ Z_v =\;1.2$. 
It is observed in the figure that the energy of the TF functional is always lower than that of KS-DFT, where the difference between the TF and the KS-DFT energy is $\sim 0.03$ a.u. at the most. Unfortunately, it is found that the inclusion of the Weizs\"{a}cker term with the factor $\lambda = \frac{1}{5}$ overcorrects the underestimation of the TF functional. Adopting the factor $\lambda = \frac{1}{9}$, which correctly reproduces the LRF of the homogeneous electron gas in the small $\bm{k}$ region, provides much better result. The fact that TFW functional with $\lambda = \frac{1}{9}$ is superior to $\lambda = \frac{1}{5}$ is somewhat surprising because it was demonstrated  in Ref. \onlinecite{Yonei1965jpsj} that the choice of $\lambda = \frac{1}{5}$ gives the smallest errors in the evaluation of the total energies of various atoms. It is possible that the result is affected by the smooth behavior of the density due to the pseudization of the atom. We also examine the performance of a sophisticated GGA (generalized gradient approximation) functional for the kinetic energy. Explicitly, we employ Eq. (12) in Ref. \onlinecite{Thakkar1992}, which was heuristically developed by integrating the several types of the GGA functionals to minimize the error for the kinetic energies of the 77 molecules. The explicit form of the functional for a spin is given by 
\begin{align}
E_{\text{kin}}^\text{GGA}[n] & = 2^{\frac{2}{3}} C_\text{TF} \int d\bm{r}\; n(\bm{r})^{\frac{5}{3}}   \notag    \\ 
\times & \left(1+\frac{A_{1} x^{2}}{1+A_{2} x \sinh ^{-1}x} -\frac{A_{3} x}{1+A_{4} x} \right)
\label{eq:GGA}
\end{align}      
where $x$ represents the inhomogeneity of the density of the spin and defined by $|\nabla n(\bm{r})|/n(\bm{r})^\frac{4}{3}$. The parameters $\{A_i\}$ in Eq. (\ref{eq:GGA}) are specified as $A_1 = 0.0055$, $A_2 = 0.0253$, $A_3 = 0.072$, and $A_4 = 2^{\frac{5}{3}}$. Despite the sophistication of the GGA functional, it offers quite minor improvement on the TFW functional with $\lambda = \frac{1}{9}$ as shown in the figure.  

Here, we discuss the contribution of the nonlocal term in Eq. (\ref{eq:Ekin_e}). As shown in the equation, the nonlocal energy is the quantity proportional to the 2nd order of the deviation of the energy electron density $\delta n^e(\epsilon) = n^e(\epsilon) - n_0^e(\epsilon)$. Thus, the contribution is zero at $Z_v = 1.0$ in principle and considered to be larger at the both ends of the plot. Actually, the third term in Eq. (\ref{eq:Ekin_e}) amounts to 0.024 a.u. and 0.013 a.u., respectively, at $Z_v = 0.8$ and $1.2$, although the contributions are much smaller as compared with those of the local term. For the calculation of Fig. \ref{Pseudo_H}, the energy coordinate $\epsilon\; (0.05 \leq \epsilon \leq 5.0)$ in the unit of a.u. is discretized by $N^e_\text{grid} = 25$ grids as listed in the \textquoteleft Supplemental Material\textquoteright. We examine the numerical robustness by setting $N^e_\text{grid} = 15$ in an additional calculation. It is demonstrated that the nonlocal terms become 0.025 a.u. at $Z_v = 0.8$, and 0.014 a.u. at $Z_v = 1.2$, which shows the numerical stability of the functional with respect to the choice of the width of the grid. 

We also examine the effect of the inclusion of the higher frequency modes in the decomposition of Eq. (\ref{eq:LRF_SD}). 
As was described at the end of Subsec. 3.D., only the term with the largest eigenvalue ($g_1 = 0.103$) is incorporated in the calculation of Fig. \ref{Pseudo_H}. It is found that considering up to the term with the secondary largest eigenvalue($g_2 = 0.016$) leads to the deviation of the kinetic energy upward by 0.01 a.u. at $Z_v = 1.2$. Furthermore, it is also revealed that including up to the third term ($g_3 = 0.00062$) worsens the kinetic energy by 0.052 a.u. Thus, the inclusion of the residual modes besides the leading term in Eq. (\ref{eq:LRF_SD}) is found to degrade the functional $E_\text{kin}^e[n]$. This trend also applies to the other end of the $Z_v$ axis. At this stage, unfortunately, we have no good reason to justify the truncation in the decomposition. However, it is speculated that some statistical noise is present in the numerical construction of the LRF due to the lack of  samplings, which may cause the creation of unphysical high frequency modes in $\chi_{0}^{e}\left(\epsilon, \epsilon^{\prime}\right)$. Actually, in our real-space grid approach, the construction of the LRF on the energy coordinate is numerically performed by projecting the data on the $64^3$ grid points in the real-space cell onto the 25 grids for the energy coordinate. Seemingly, the amount of the data assigned to an energy coordinate $\epsilon$ is not sufficient when the coordinate corresponds to the atomic core region in particular. To increase the sampling points, one may consider to apply the double-grid technique\cite{rf:ono1999prl} to the core region. We found, however, that the straightforward implementation of the method leads to the destruction of the crucial condition that the matrix $\chi_{0}^{e}\left(\epsilon, \epsilon^{\prime}\right)$ is positive semi-definite. In the following, we provide the results for the remaining pseudo atoms, i.e., He, Ne, and Ar. 

\begin{figure}[h]
\centering
\scalebox{0.35}[0.35] {\includegraphics[trim=0 5 10 10,clip, ]{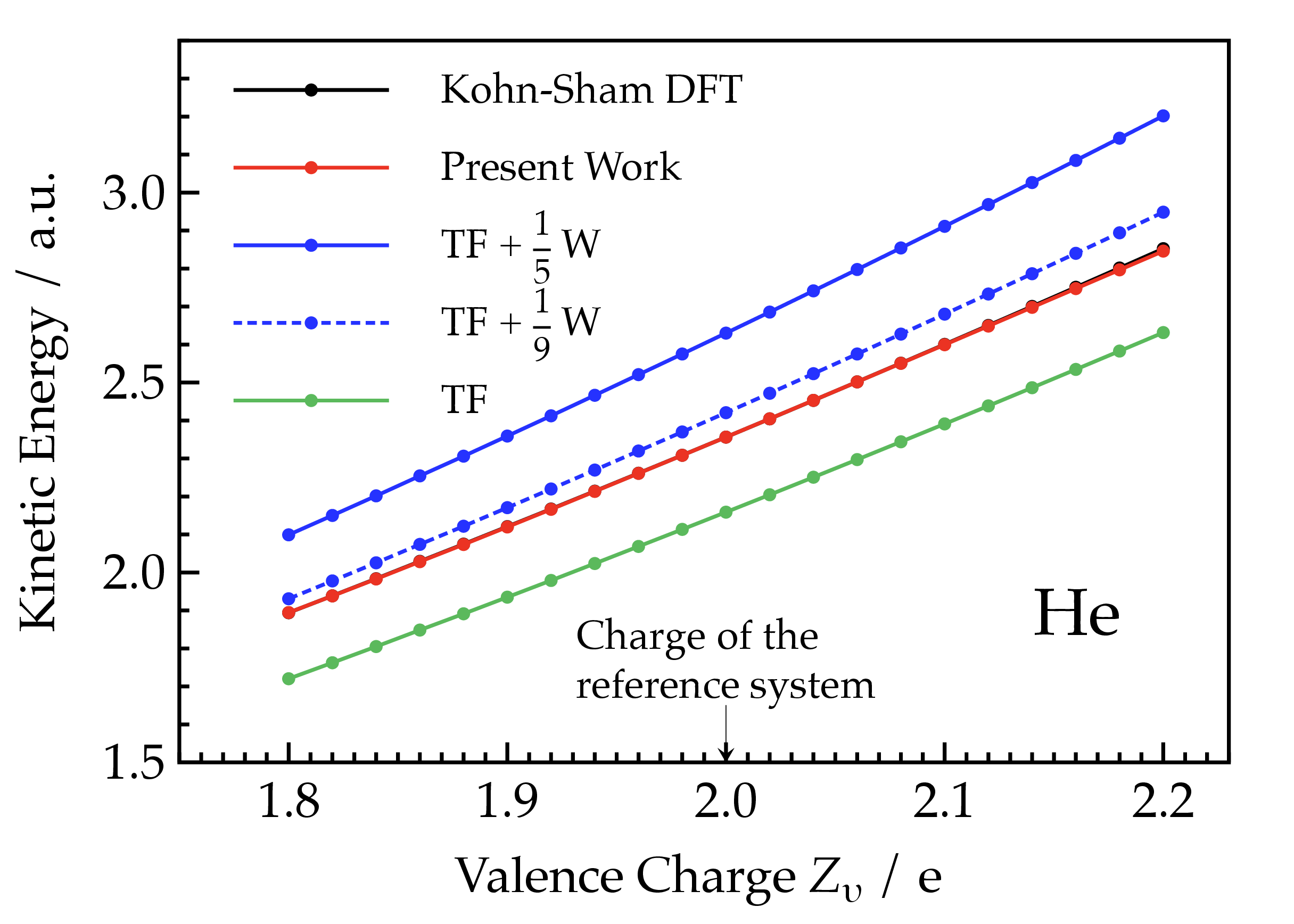}}            
\caption{\label{Pseudo_He}  Plots of the kinetic energies computed by various kinetic functionals for the variation of the valence charge $Z_v$ of the pseudo He atom. The atom with $Z_v = 2.0$ is taken as the reference system. The plots for \textquoteleft Kohn-Sham DFT\textquoteright\; and \textquoteleft Present Work\textquoteright\; are almost indiscernible in the figure. The other notations are the same as in Fig. \ref{Pseudo_H}. }
\end{figure}

In Fig. \ref{Pseudo_He} the profiles of the kinetic energies for the pseudo He atom are presented. The energy range of the plots is extended by about 3 times as compared with that for H atom because the depth of the local pseudopotential for He is much deeper than that for H as shown in Fig. \ref{VPS} and the number of electrons is twice that of H. It is clearly recognized in the figure that the plots have quite similar trend as those for the pseudo H atom.  It is striking that the kinetic energy given by Eq. (\ref{eq:Ekin_e}) excellently agrees with that by KS-DFT over the whole range of $Z_v$. Actually, the two plots are almost indistinguishable in the graph. We note, however, that absolute difference in the kinetic energy at $Z_v = 2.2$ between the present work and the KS-DFT is evaluated as $0.006$ a.u., and this is found to be about the half of the difference between them for the pseudo H atom at $Z_v = 0.8$.        

\begin{figure}[h]
\centering
\scalebox{0.35}[0.35] {\includegraphics[trim=0 5 10 -5,clip, ]{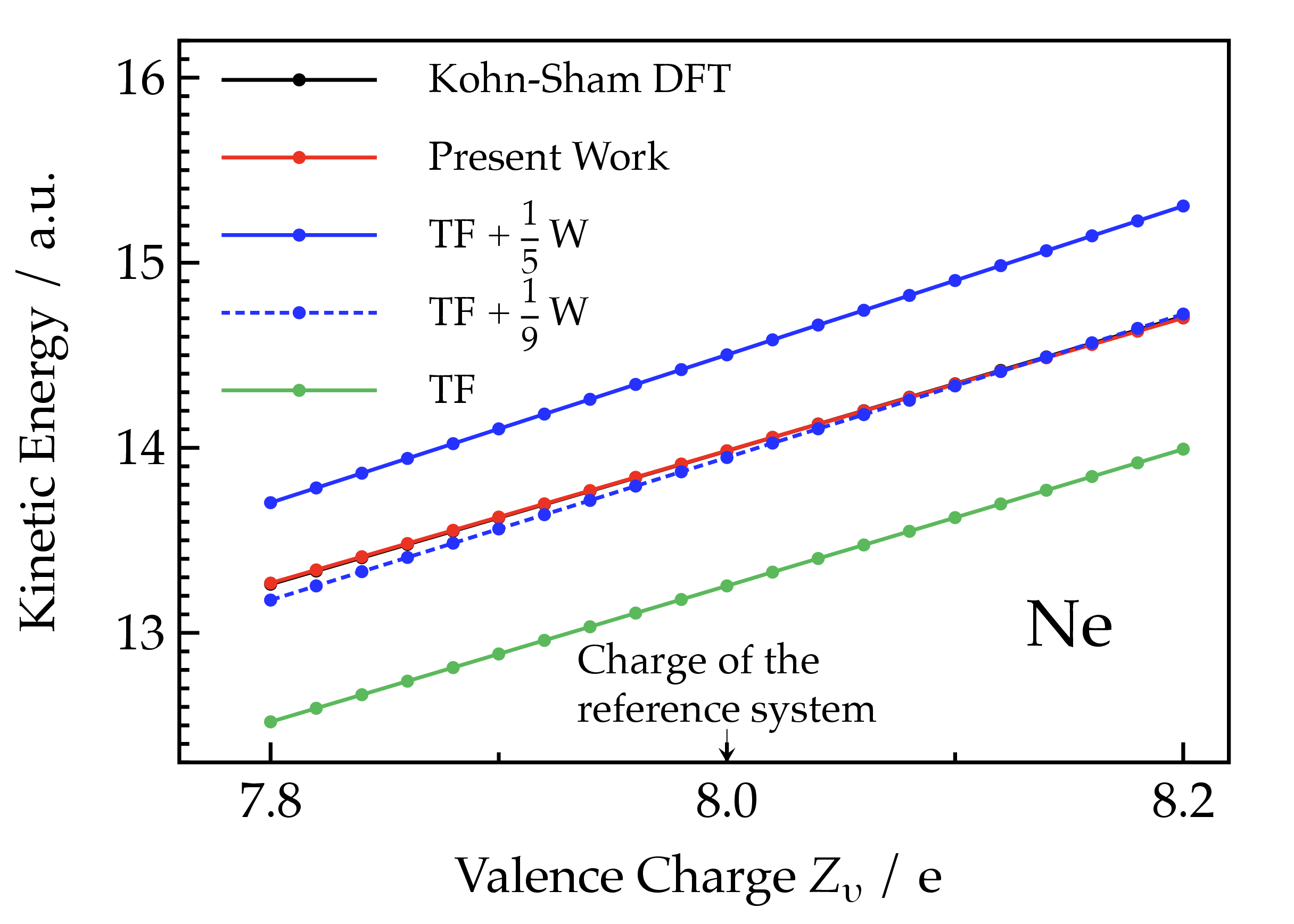}}            
\caption{\label{Pseudo_Ne}  Plots of the kinetic energies computed by various kinetic functionals for the variation of the valence charge $Z_v$ of the pseudo Ne atom. The atom with $Z_v = 8.0$ is taken as the reference system. The plots for \textquoteleft Kohn-Sham DFT\textquoteright\; and \textquoteleft Present Work\textquoteright\; are almost indiscernible in the figure. The other notations are the same as in Fig. \ref{Pseudo_H}.  }
\end{figure}

In Fig. \ref{Pseudo_Ne}  the results for the pseudo Ne atom are plotted. The range of the energy variation is 1.5 times as large as that of the He atom. This is attributed to the fact that the number of the valence electrons for Ne is 4 times as many as that for He. Note, however, that the valence electrons in Ne reside outside the core region due to the repulsive potential $\Delta V_{\mathrm{s}}(r)$ augmented in the local pseudopotential drawn in Fig. \ref{VPS}. Again, it is found that the coincidence between the present work and the KS-DFT is quite excellent. The maximum absolute deviation from the KS-DFT value is $0.006$ a.u. that occurs at $Z_v = 7.8$. As a notable feature specific to the results for Ne, the Thomas-Fermi functional corrected with 1/9 Weizs\"{a}cker term shows good agreements with the KS-DFT in the range of $ 8.0 \leq Z_v \leq 8.2$.    

\begin{figure}[h]
\centering
\scalebox{0.35}[0.35] {\includegraphics[trim=0 5 10 -5,clip, ]{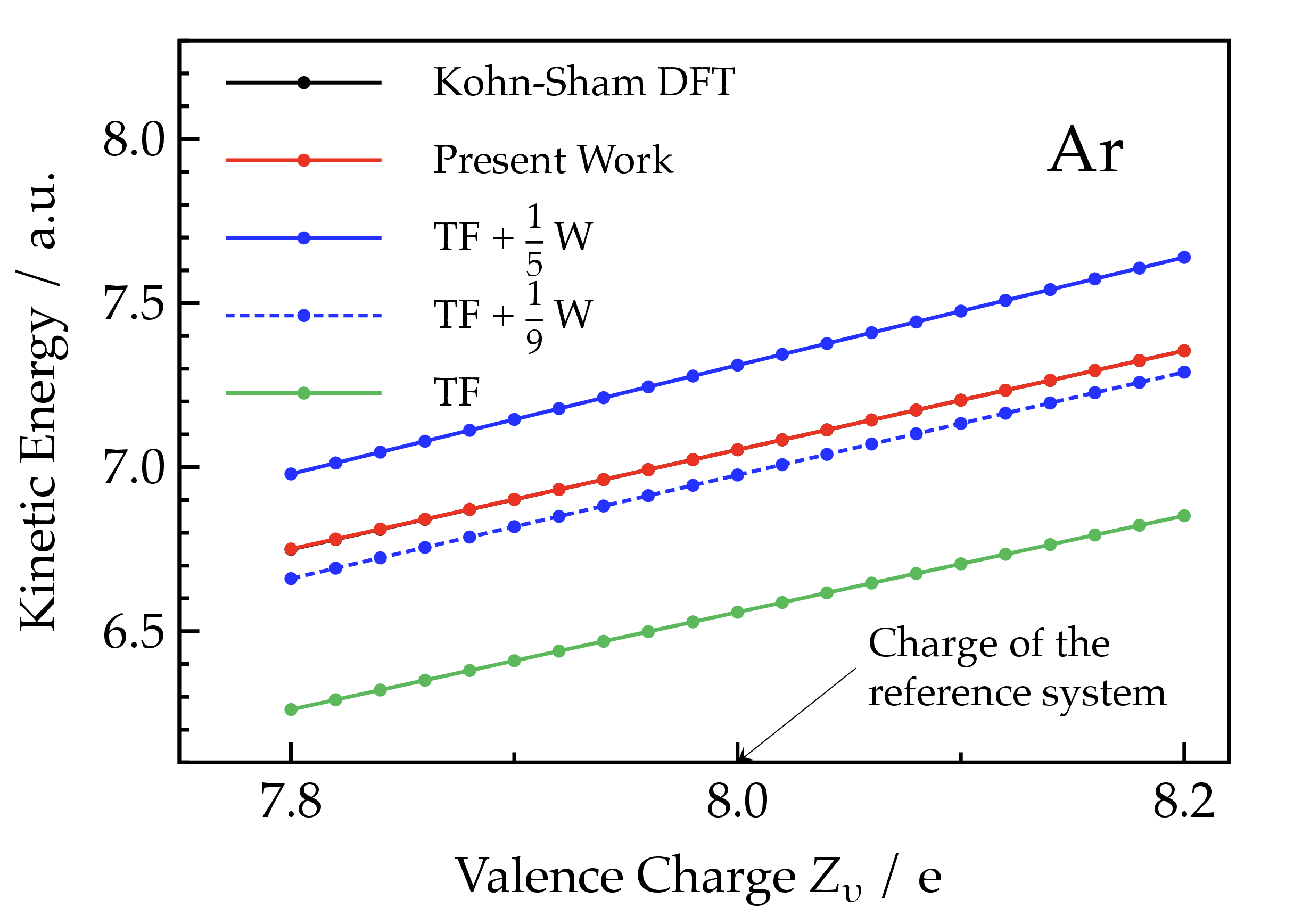}}            
\caption{\label{Pseudo_Ar}  Plots of the kinetic energies computed by various kinetic functionals for the variation of the valence charge $Z_v$ of the pseudo Ar atom. The atom with $Z_v = 8.0$ is taken as the reference system. The plots for \textquoteleft Kohn-Sham DFT\textquoteright\; and \textquoteleft Present Work\textquoteright\; are almost indiscernible in the figure. The other notations are the same as in Fig. \ref{Pseudo_H}.  }
\end{figure}

Lastly, we provide the results for the pseudo Ar atom in Fig. \ref{Pseudo_Ar}. In the figure, it is shown that the range of the kinetic energy variation is reduced to about the half of the energy range for Ne. This can be attributed to the facts that the local pseudopotential of Ar (see Fig. \ref{VPS}) is much shallower than that of Ne, and the potential valley of Ar is located farther from the atomic core than that of Ne. It is worthy of noting that the present work shows the best agreements with the KS-DFT values in the application to Ar among the other atoms. Actually, the maximum deviation in the energy from the KS-DFT values is found to be only $\sim$0.001 a.u. However, this is simply because the nonlocal energy gives the minor contribution to the kinetic energy in the Ar system. Explicitly, the nonlocal energy in the functional $E_\text{kin}^e$ of Eq. (\ref{eq:Ekin_e}) is evaluated to be 0.007 a.u. at $Z_v = 8.2$.  

We close this subsection by making a brief remark on the kinetic energy functional $E_\text{kin}^e$ using a non self-consistent density. First, it was demonstrated that the functional is able to compute the kinetic energies with reasonable accuracies as compared with those obtained by the Kohn-Sham DFT. Second, it was found that the nonlocal term, represented with a functional of the energy electron density $n^e(\epsilon)$, certainly improves the description of the kinetic energy although the contribution to $E_\text{kin}^e$  is proportional to the second order with respect to $\delta n^e$. In the next subsection, we provide the self-consistent densities obtained through the present OF-DFT approach. 
\subsection{Self-consistent Field Calculations with the OF-DFT Method}
\begin{figure}[h]
\centering
\scalebox{0.35}[0.35] {\includegraphics[trim=0 5 10 -5,clip, ]{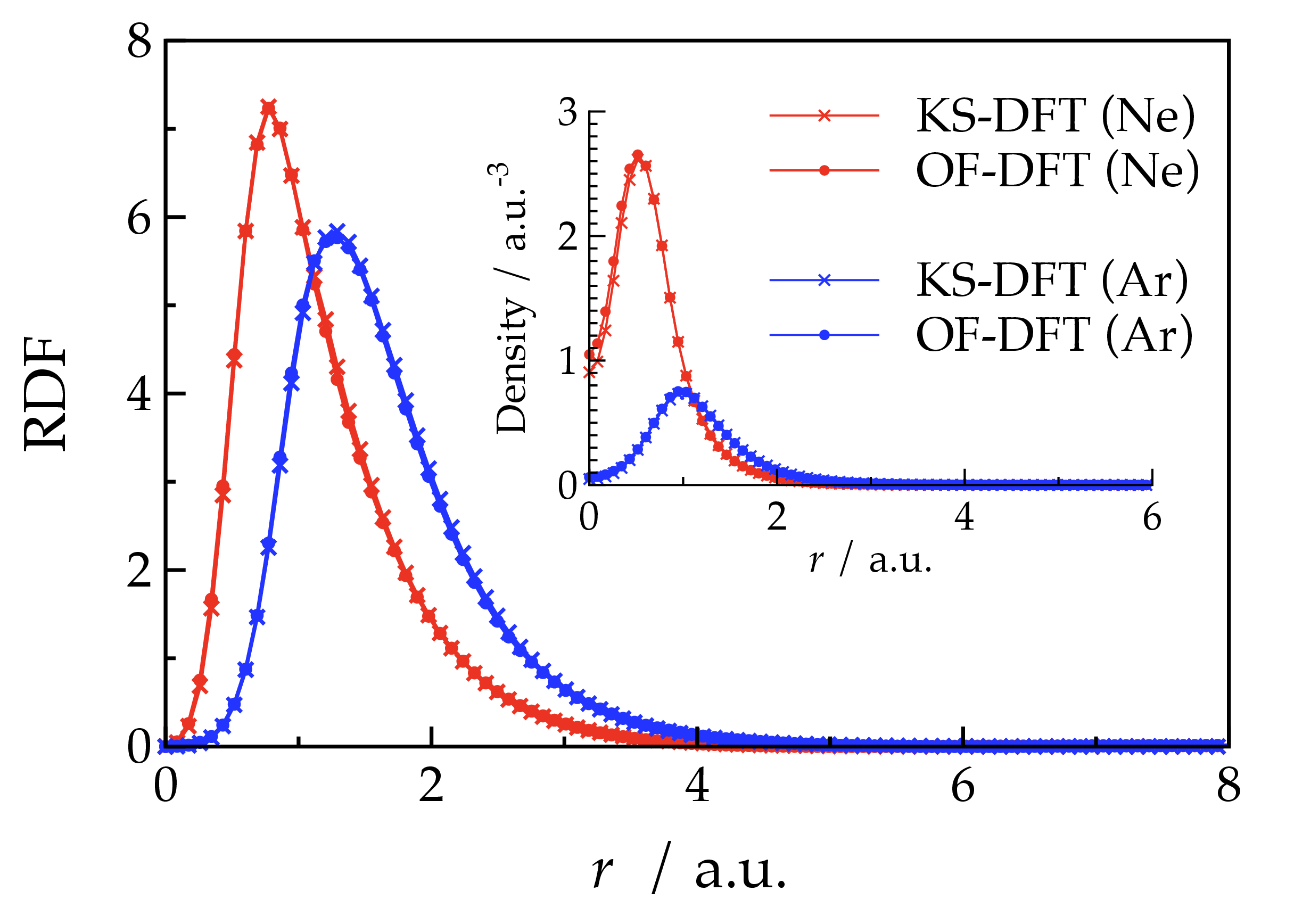}}            
\caption{\label{RDF}  Plots of the radial distribution functions (RDFs) for the electrons in the pseudo Ne and Ar atoms. Superposed in the figure is the corresponding density profiles along the radial distance $r$. The electron densities are obtained through the SCF calculations. The valence charges $Z_v$ of these systems are set at $Z_v = 8.2$ in the unit of the elementary charge.  }
\end{figure}

In this subsection, the radial distribution functions (RDFs) for electrons in the pseudo Ne and Ar atoms with $Z_v = 8.2$ are presented, where the corresponding electron densities are optimized through the self-consistent field (SCF) calculations utilizing Eqs. (\ref{eq:SCF_e}) and (\ref{eq:tilde_chi}). The parameter $\eta$ in Eq. (\ref{eq:SCF_e}) is set at $0.05$. The convergence thresholds are $5.0\times10^{-7}$ a.u. for the total energy, and $5.0\times10^{-9}$ a.u.$^{-3}$ for the density. The initial guess for the density for each system is the electron density $n_0$ of the system with $Z_v = 8.0$. The 4-points polynomial interpolation method is utilized to increase the sampling points for the construction of the smooth RDFs. Explicitly, $(100, 200, 400)$ grid points are yielded for the spherical coordinates $(r, \theta, \phi)$ and the density on each grid is evaluated by the polynomial interpolation of the rectangular grids, where the range of the radial distance $r$ is $0 \leq r \leq 8.604$ a.u. The results are plotted in Fig. \ref{RDF}. We found in the both plots that the RDFs optimized by the OF-DFT reasonably agree with those given by KS-DFT. We note, however, that the kinetic energies $E_\text{kin}$ by the OF-DFT are somewhat different from those by KS-DFT calculations. Explicitly, $E_\text{kin}$ for the Ne atom is evaluated as $14.685$ a.u. by the OF-DFT, while the KS-DFT gives $E_\text{kin} = 14.709$ a.u. For the Ar atom, $E_\text{kin}$ are computed as $7.344$ and $7.355$ a.u., by the OF-DFT and KS-DFT methods, respectively. Thus, it was found the kinetic energy as well as the density are degraded through the SCF procedure in the OF-DFT calculation. 

We should make additional remarks on the construction of the RDF for the Ne atom. Specifically in the SCF calculation for the Ne atom, we actually adopt the BHS local potential $V_{\text{loc}}^{\text{BHS}}(r)$ of the Ar atom\cite{rf:bachelet1982prb} as the potential $\upsilon_\text{def}(\bm{r})$ to create the energy coordinate $\epsilon$. The use of $V_{\text{loc}}^{\text{BHS}}(r)$ for Ne as $\upsilon_\text{def}(\bm{r})$ necessitates the wide energy range for the energy coordinate because the Ne potential is very steep around the atomic core. Therefore, in the construction of the response function using such a defining potential, a lot of bins become null in the region with large energy coordinate due to the low resolution of the real-space grids around the core. This might give rise to a numerical instability in the SCF calculation.  In principle, the potential of any atomic core can be indeed used as a defining potential for another atom within the framework of the DFT based on the energy electron density\cite{Takahashi2018jpb} since the potential due to the nucleus is always spherically symmetric. Actually, the use of the Ar potential rather than the Ne potential in computing the plot in Fig. \ref{Pseudo_Ne} makes a difference of only $\sim 0.5$ kcal/mol ($\sim8\times10^{-4}$ a.u.) in the kinetic energy at $Z_v = 8.2$. 

We also note an unfavorable phenomenon found in the SCF calculation for the Ne atom. That is, the electron population on the atomic core increases gradually through the SCF iterations, which seriously affects the density convergence. This suggests that the nonlocal term in the kinetic potential does not work appropriately on the atomic core region since only the nonlocal potential is dependent on the electron density as seen in Eq. (\ref{eq:v_kin_e}). The drawback might be attributed to the poor resolution of the grids around the core. \textcolor{black}{The problem also emerges when $V_{\text{loc}}^{\text{BHS}}(r)$ for Ne is used as $\upsilon_\text{def}(\bm{r})$ because it also makes no contribution to increase the resolution around the core.}  Here, we apply an ad hoc approach to compensate for the error. Explicitly, we mix the kinetic potential with the Thomas-Fermi (TF) potential only on the core region using a Gaussian weight function $F(r)$. The explicit form of the kinetic potential $\overline{\upsilon}_{\text {kin }}^{e}$ with the patch on the core is given by
\begin{align}
\overline{\upsilon}_{\text {kin }}^{e}[n](\bm{r}) = & \upsilon_{\text {kin}}\left[n_{0}\right](\bm{r}) +F(r) \upsilon_{\text {kin }}^{\text {TF}}[n](\bm{r})     \notag    \\
&+(1-F(r)) \upsilon_{\text {kin}}^\text{nloc}\left[n^{e}\right](\epsilon)
\label{eq:patch}
\end{align}
where the potential $\upsilon_{\text {kin}}^\text{nloc}$ is the second term of the right hand side of Eq. (\ref{eq:v_kin_e}) and the weight function $F(r)$ is defined as
\begin{equation}
F(r) = \exp(-\zeta r^2)    \;\;\;\;\;\; (\zeta > 0)
\label{eq:weight}
\end{equation}
In Eq. (\ref{eq:patch}), the potential $\upsilon_{\text {kin }}^{\text {TF}}[n](\bm{r})$ is expressed in terms of the TF potential, thus, 
\begin{equation}
\upsilon_\text{kin}^\text{TF}[n](\bm{r})=\left.\frac{\delta \upsilon_\text{TF}[n](\bm{r})}{\delta n(\bm{r})}\right|_{n_{0}}\left(n(\bm{r})-n_{0}(\bm{r})\right)
\end{equation}
where the potential $\upsilon_\text{TF}$ is given by the functional derivative of Eq. (\ref{eq:E_TF}), that is, $\delta E_\text{TF}[n]/\delta n(\bm{r})$. Notice that the potential $\upsilon_\text{kin}^\text{TF}[n](\bm{r})$ is designed so that it becomes zero when the density $n(\bm{r})$ coincides with $n_0(\bm{r})$.  The parameter $\zeta$ in Eq. (\ref{eq:weight}) specifies the width of the Gaussian and it is set at $\zeta = 5.0$ a.u.$^{-2}$ in the present calculation. With this choice of $\zeta$, the weight $F(r)$ is reduced to less than 0.2 within the distance $r=2h$ for the grid width of $h=0.2868$ a.u. Thus, the potential $\upsilon_\text{kin}^\text{TF}$ in Eq. (\ref{eq:patch}) applies only to the limited region around the core. It was demonstrated that the treatment improves the SCF convergence in the electron density. It was also found in the calculation of $Z_v = 8.2$ that the contribution of the TF energy to the kinetic energy amounts to $5.2$ kcal/mol after the SCF convergence.      

\section{Conclusion and Perspective}
In this article, we developed a kinetic energy functional $E_\text{kin}^e[n]$ within the framework of the density-functional theory based on the energy electron density $n^e(\epsilon)$\cite{Takahashi2018jpb} for the purpose of realizing the orbital-free DFT(OF-DFT) calculations. The functional includes the nonlocal term described with a linear-response function(LRF) of a reference system as was introduced in the Wang-Teter functional\cite{Wang1992prb}. As a notable feature, the LRF, which constitutes the integral kernel of the nonlocal term, is represented on the energy coordinate $\epsilon$, thus, $\chi_0^e(\epsilon, \epsilon^\prime)$. In addition, an atomic system is taken as a reference system in contrast to the conventional approach that utilizes LRF of the homogeneous electron gas. The kinetic functional $E_\text{kin}^e[n]$ was formulated by means of the coupling-parameter integration scheme. The functional was applied to the calculation of the kinetic energies of the pseudo atoms which stem from H, He, Ne, and Ar atoms. Explicitly, the kinetic energy of each atom was computed with respect to the variation of the valence charge $Z_v$, where the atom with its original charge $Z_v$ was adopted as the reference system to construct $\chi_0^e(\epsilon, \epsilon^\prime)$. It was demonstrated that the functional $E_\text{kin}^e[n]$ is able to provide the kinetic energies in good agreements with those obtained by the Kohn-Sham DFT(KS-DFT) for the given non self-consistent electron densities $n$.   

We also devised a scheme to perform the self-consistent field (SCF) calculation using $E_\text{kin}^e[n]$ although the sophistications to expedite the convergence were not made. The radial distribution functions (RDFs) of the electrons in the pseudo Ne and Ar with $Z_v = 8.2$ were obtained through the SCF calculations. The RDFs by the OF-DFT calculations showed reasonable agreements with those given by KS-DFT calculations. Thus, the reliability of the method was demonstrated. 

It should be noticed that the application of the present approach to an actual system of interest such as molecules is inefficient since it necessitates the individual solution of the KS equation for a certain reference system to build the LRF for each system.  Our strategy in the near future to solve the problem is to compute and store beforehand the LRFs of the constituent atomic systems. Then, we take the overlap of the component LRFs on the energy coordinate to construct the total LRF of the target molecule. We note that the construction of the molecular LRF on the energy coordinate is justified by the DFT based on the energy electron density\cite{Takahashi2018jpb}. Although the projection of the LRF onto the energy coordinate itself is time consuming, the computational cost is proportional to the system size.  The reference electron density $n_0$ will also be constructed from the overlap of the electron densities of the constituent atoms. The corresponding kinetic energy and the potential for the reference system will be evaluated by some existing GGA functional for instance. These methods will be developed and examined in a forthcoming issue.  
\\
\\
\\
\\
\\    
 
\begin{acknowledgments}
The author is grateful to Prof. N. Matubayasi in Osaka university for the fruitful discussions on the pseudo inverse method utilized for the inversion of the linear-response matrix. This paper was supported by the Grant-in-Aid for Scientific Research on Innovative Areas (No. 23118701) from the Ministry of Education, Culture, Sports, Science, and Technology (MEXT); the Grant-in-Aid for Challenging Exploratory Research (No. 25620004) from the Japan Society for the Promotion of Science (JSPS); and the Grant-in-Aid for Scientific Research(C) (No. 17K05138) from the Japan Society for the Promotion of Science (JSPS). This research also used computational resources of the HPCI system provided by Kyoto, Nagoya, and Osaka university through the HPCI System Research Project (Project IDs: hp170046, hp180030, hp180032, hp190011, and hp200016).
\end{acknowledgments}  

\appendix

\section{Linear-response Function}
Here, we prove that the linear-response function (LRF) $\chi_0^e(\epsilon, \epsilon^\prime)$ in Eq. (\ref{eq:chi_e}) defined on the energy coordinate is positive semi-definite. We start the discussion with the LRF $\chi_{0}\left(\bm{r}, \bm{r}^{\prime}\right)$  in Eq. (\ref{eq:PT2}) defined on the spatial coordinate. First, we provide the proof that $\chi_{0}\left(\bm{r}, \bm{r}^{\prime}\right)$ is positive semi-definite. To this end, we consider a quantity $Q$ with respect to an arbitrary function $q(\bm{r})$,
\begin{equation}
Q=\int d \bm{r} d \bm{r}^{\prime} q(\bm{r}) \chi_{0}\left(\bm{r}, \bm{r}^{\prime}\right) q^{*}\left(\bm{r}^{\prime}\right)
\label{eq:Q}
\end{equation} 
\\
For our purpose, it is sufficient to prove that the relation $Q \geq 0$ always holds. By substituting Eq. (\ref{eq:PT2}) into Eq. (\ref{eq:Q}), one readily obtains, 
\begin{align}
Q &= \int d\bm{r} d\bm{r}^\prime  \sum_{i}^{\text {occ}} \sum_{a}^{\text {vir}} \frac{1}{\varepsilon_{a}^{0}-\varepsilon_{i}^{0}}q(\bm{r}) \varphi_{i}^{0*}(\bm{r}) \varphi_{a}^{0}(\bm{r})      \notag           \\
& \;\;\;\;\;\;\;\;  \times \varphi_{a}^{0*}(\bm{r}^\prime) \varphi_{i}^{0}(\bm{r}^\prime) q^{*}(\bm{r}^\prime)         \notag     \\
& =\sum_{i}^\text{occ} \sum_{a}^\text{vir} \frac{1}{\epsilon_{a}^{0}-\epsilon_{i}^{0}} \int d \bm{r} d \bm{r}^{\prime} \Phi_{i a}(\bm{r}) \Phi_{i a}^{*}\left(\bm{r}^{\prime}\right)     \notag   \\
& =\sum_{i}^\text{occ} \sum_{a}^\text{vir} \frac{1}{\epsilon_{a}^{0}-\epsilon_{i}^{0}}\left|S_{i a}\right|^{2}   
\label{eq:Q2}
\end{align}
where the function $\Phi_{i a}(\bm{r})$ in the right hand side of the second equality is defined as $\Phi_{i a}(\bm{r}) = q(\bm{r}) \varphi_{i}^{0*}(\bm{r}) \varphi_{a}^{0}(\bm{r})$ and $S_{ia}$ in the last equality is given by $S_{ia} = \int d\bm{r} \Phi_{i a}(\bm{r})$. Note that the quantity $Q$ becomes zero when the function $q(\bm{r})$ is constant. Thus, it is proved that $\chi_{0}\left(\bm{r}, \bm{r}^{\prime}\right)$ is positive semi-definite. Then, the it is rather straightforward to prove that $\chi_0^e(\epsilon, \epsilon^\prime)$ is also positive semi-definite.  For an arbitrary function $q^e(\epsilon)$, it reads, 
\begin{align}
\int & d \epsilon d \epsilon^{\prime} q^{e}(\epsilon)  \chi_{0}^{e}\left(\epsilon, \epsilon^{\prime}\right) q^{e}\left(\epsilon^{\prime}\right)   \notag \\
 = & \int d \epsilon d \epsilon^{\prime} q^{e}(\epsilon) q^{e}\left(\epsilon^{\prime}\right)     \notag   \\
&\;\;\;\;\;\; \times \int d \bm{r} d \bm{r}^{\prime} \chi_{0}\left(\bm{r}, \bm{r}^{\prime}\right) \delta(\epsilon-v(\bm{r})) \delta\left(\epsilon^\prime -v\left(\bm{r}^{\prime}\right)\right)     \notag    \\
= & \int d \bm{r} d \bm{r}^{\prime} \chi_{0}\left(\bm{r}, \bm{r}^{\prime}\right) \int d \epsilon\; q^{e}(\epsilon) \delta(\epsilon-v(\bm{r}))  \notag      \\
&\;\;\;\;\;\; \times   \int d \epsilon^{\prime} q^{e}\left(\epsilon^{\prime}\right) \delta\left(\epsilon^{\prime}-v\left(\bm{r}^{\prime}\right)\right)   \notag \\
= &  \int d \bm{r} d \bm{r}^{\prime}\; \theta(\bm{r}) \chi_{0}\left(\bm{r}, \bm{r}^{\prime}\right) \theta(\bm{r}^\prime)    \geq 0
\end{align}
where $\theta(\bm{r})$ is defined as $\int d\epsilon q^{e}(\epsilon) \delta(\epsilon-v(\bm{r}))$. Thus, $\chi_0^e(\epsilon, \epsilon^\prime)$ is also positive semi-definite. Hence,  $\chi_0^e(\epsilon, \epsilon^\prime)$ can be inverted through a manipulation such as a pseudo inverse method. 

\section{Grid Error in Response Function}
Since we utilize the real-space grid (RSG) approach in representing the response function $\chi_0(\bm{r},\bm{r}^\prime)$, its projection $\chi_0^e(\epsilon,\epsilon^\prime)$ onto the energy energy coordinate $\epsilon$ (Eq. (\ref{eq:chi_e})) is affected by the grid error inherent in $\chi_0(\bm{r},\bm{r}^\prime)$. In this Appendix, we demonstrate that the terms with small eigenvalues in the spectral decomposition in Eq. (\ref{eq:LRF_SD}) are comparable in magnitude to the grid error in $\chi_0^e(\epsilon,\epsilon^\prime)$. To this end, we employ the double-grid technique. Explicitly, the values $\chi_0(\bm{r},\bm{r}^\prime)$ at the double grids (DGs) are evaluated through numerical interpolations of the original coarse grids (CGs) to save the computer resources. In the next paragraph, a concise explanation is provided for the DG method applied to the response function of the Ar atom. 

Hereafter, it is assumed that the origin of the position vectors is set at the atomic center. Since $\chi_0(\bm{r},\bm{r}^\prime)$ as a function of $\bm{r}$ has a cylindrical symmetry around the axis $Z$ being oriented to the vector $\bm{r}^\prime$, the original data $\chi_0(\bm{r},\bm{r}^\prime)$ at CGs, which serve to evaluate the response function at DGs, can be reduced to three-dimensional data $\widetilde{\chi}_0$ without losing the information contents. Explicitly, the reference response function $\widetilde{\chi}_0$ has the three arguments, thus $\widetilde{\chi}_0(d; Y, Z)$, where $d$ represents the length of the vector $\bm{r}^\prime$, and the coordinates $(Y, Z)$ specify the position $\bm{r}$ on the plane which includes $\bm{r}$ and the axis $Z$. We, thus evaluate $\widetilde{\chi}_0(d; Y, Z)$ beforehand at CGs  with the grid width $h$ given in subsection III A. The DGs are then placed within each CG at the interval of $h/N_\text{DG}$ along each axis ($1 \leq N_\text{DG} \leq 4 $, $N_\text{DG} \in \mathbb{Z}$). Let $\bm{r}_i$ and $\bm{r}_j$ be the position vectors of the DGs with indices $i$ and $j$, respectively, the corresponding response function $\chi_0(\bm{r}_i,\bm{r}_j)$ is to be obtained through the following procedure. Based on the above discussion, the six-dimensional coordinate $(\bm{r}_i,\bm{r}_j)$ can be readily cast into the reduced variables $(d; Y, Z)$, thus
\begin{equation}
\left\{\begin{array}{l}
d = |\bm{r}_j| \\
Y = |\bm{r}_i| \sin \theta \\
Z = |\bm{r}_i| \cos \theta
\end{array}\right.  
\label{eq:B1}
\end{equation}
where $\theta$ represents the angle between the vectors $\bm{r}_i$ and $\bm{r}_j$, and defined by 
$\cos \theta = \frac{\bm{r}_i \cdot \bm{r}_j}{|\bm{r}_i| \cdot |\bm{r}_j|}$. We note that the indices $i$ and $j$ in Eq. (\ref{eq:B1}) are interchangeable because of the symmetry of the function $\chi_0(\bm{r}_i,\bm{r}_j)$. In this work, the response function $\chi_0(\bm{r}_i,\bm{r}_j)$ between the coordinates $\bm{r}_i$ and $\bm{r}_j$ at the DGs is evaluated through the 4th-order polynomial interpolation of the  function $\widetilde{\chi}_0(d; Y, Z)$ defined at CGs. The DGs are placed within the CGs of which distances from the atomic core are less than $5.0$ a.u. The numerical detail of the interpolation is presented in Ref. \onlinecite{takahashi2001jpca}.  

\begin{figure}[h]
\centering
\scalebox{0.35}[0.35] {\includegraphics[trim=0 7 10 -4,clip, ]{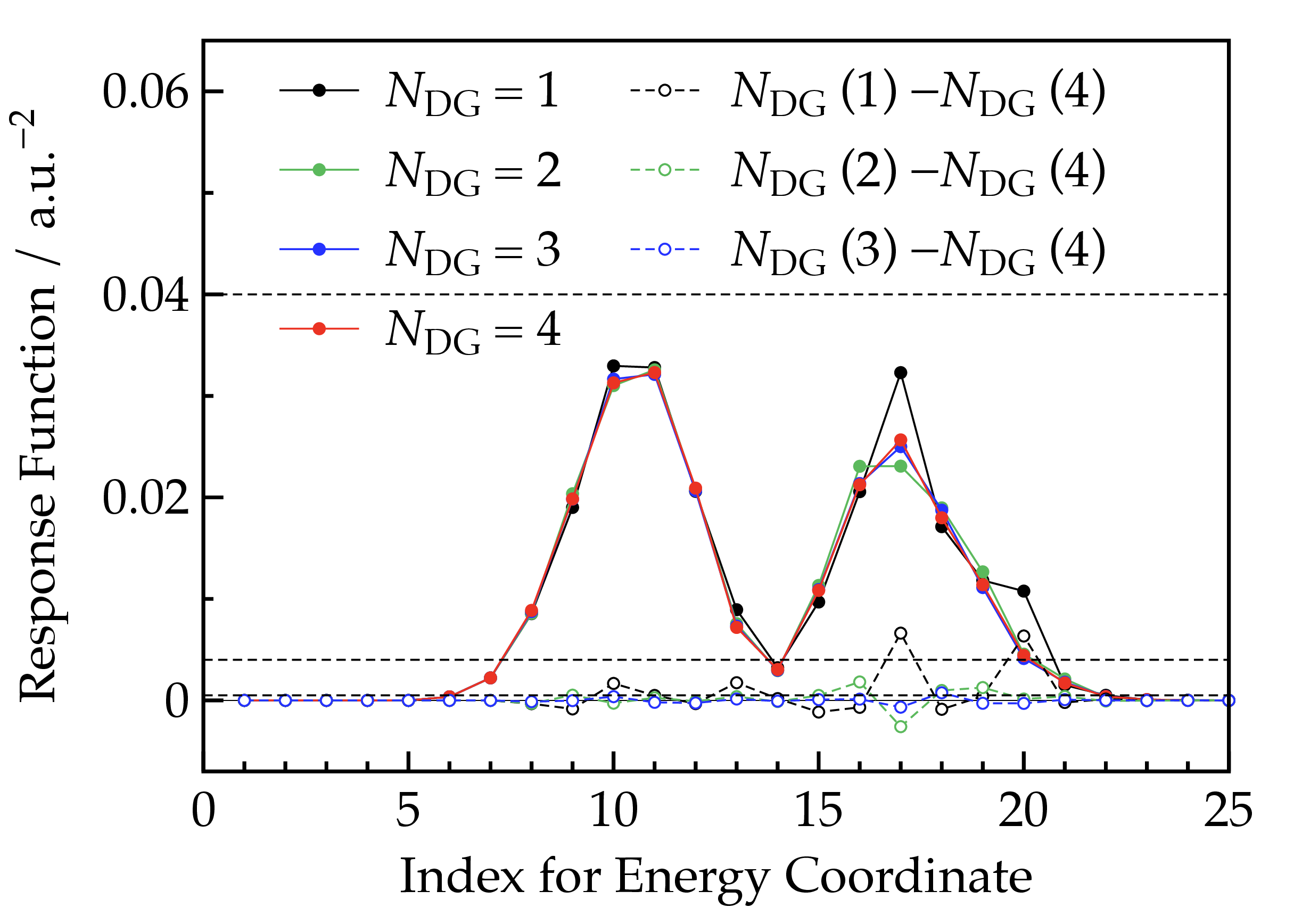}}            
\caption{\label{Grid_Error}  The diagonal elements of the linear-response function $\chi_0^e(\epsilon, \epsilon^\prime)$ for the pseudo Ar atom as functions of the grid index for the energy coordinate $\epsilon$. The real lines are the plots for the calculations with $N_\text{DG}$. The plots with the broken lines are the results for $N_\text{DG} = 1, 2, $ and 3 subtracted by that for $N_\text{DG} = 4$. The three horizontal broken lines represent the values, 0.04, 0.004, and 0.0005, which are approximately the maximum values of the component response functions $g_i \gamma_i(\epsilon)\gamma_i(\epsilon^\prime)$ with the eigenvalues $g_1$, $g_2$, and $g_3$ shown in Figs. \ref{LRF}(b), (c), and (d), respectively. }
\end{figure}

In Fig. \ref{Grid_Error} the diagonal elements of the response functions $\chi_0^e(\epsilon, \epsilon^\prime)$ as functions of the index for the discrete energy coordinate $\epsilon$ are presented for the various $N_\text{DG}$. It is seen in the figure that all the graphs are characterized by two distinct peaks although they differ slightly each other due to the difference in the width of the DG. It is also recognized that the graphs reasonably converge as the grid interval decreases, which shows that the calculation with $N_\text{DG} = 4$ offers the smallest grid error among the plots as expected. We subtract the result for $N_\text{DG} = 4$ from that for other $N_\text{DG}$ to estimate the effect of the grid width on the response function. Since the result for $N_\text{DG} = 1$ is equivalent in principle to the calculation done in the main text, the graph \textquoteleft $N_\text{DG}(1)-N_\text{DG}(4)$\textquoteright\; approximately exhibits the grid error in representing $\chi_0^e(\epsilon, \epsilon^\prime)$ in the present work. The three horizontal broken lines in the figure approximately show maximum values of the plots shown in Figs. \ref{LRF}(b), (c), and (d). It is observed that the grid error is comparable or even larger in magnitude than the response functions with $g_2$ and $g_3$. Therefore, the high-frequency modes $\gamma_i$ in the response function are seriously contaminated by artifacts due to the grid errors. These contributions should be excluded from the summation of Eq. (\ref{eq:LRF_inv}) because the contamination is erroneously amplified by the multiplication of $g_i^{-1}$. Seemingly, it is quite difficult to extract the grid errors from the eigenvectors. One might consider that the DG method will alleviate the problem. We found, however, that the introduction of the DGs destroys unfortunately the crucial condition that $\chi_0^e(\epsilon, \epsilon^\prime)$ is positive semi definite. This problem could be attributed to the error in the polynomial interpolation utilized in the DG method.  

Although the truncation in the expansion of Eq. (\ref{eq:LRF_SD}) within a few terms would not seriously affect the accuracy of the kinetic energy, it is desirable to make some devices in the future work to alleviate the problem. First, some approximate method should be developed to estimate the error in the response function in terms of the grid size in the real-space cell, which will offer a method of rational truncations. Second, a new numerical device for the DG approach should be made to maintain the positive semi definiteness in the construction of the matrix $\chi_0^e(\epsilon, \epsilon^\prime)$. These subjects will be addressed in the forthcoming issues.                 

\providecommand{\noopsort}[1]{}\providecommand{\singleletter}[1]{#1}%

\end{document}